\documentclass[pra,twocolumn,preprintnumbers,amsmath,amssymb]{revtex4}

\usepackage{graphicx}
\usepackage{dcolumn}
\usepackage{bm}
\usepackage{color}
\usepackage{epstopdf}
\usepackage[normalem]{ulem}
\usepackage{cancel}

\usepackage{ulem}

\newcommand{\br}{ {\bm r}}

\def\psih{{\hat{\psi}}}
\def\psihd{{\hat{\psi}^\dagger}}
\def\phih{{\hat{\phi}}}
\def\phihd{{\hat{\phi}^\dagger}}

\def\br{{\bm r}}

\begin{document}



\title{Spatio-temporal equilibrium thermodynamics of guided optical waves\\ 
at positive and negative temperatures}
\author{Lucas Zanaglia$^1$, Josselin Garnier$^{2}$,  Claire Michel$^{1,3}$, Val\'erie Doya$^{1}$, Mario Ferraro$^{4}$, Stefan Wabnitz$^{5}$, Iacopo Carusotto$^6$, Antonio Picozzi$^{7}$}
\affiliation{$^{1}$ Universit\'e C\^ote d'Azur, CNRS, Institut de Physique de Nice, Nice, France}
\affiliation{$^{2}$ CMAP, CNRS, Ecole polytechnique, Institut Polytechnique de Paris, 91120 Palaiseau, France}
\affiliation{$^{3}$Institut Universitaire de France (IUF), 1 rue Descartes, 75005 Paris, France}
\affiliation{$^{4}$Department of Physics, University of Calabria, Via Pietro Bucci, Rende, 87036, (CS), Italy}
\affiliation{$^{5}$Department of Information Engineering, Electronics, and Telecommunications, Sapienza University of Rome, Via Eudossiana 18, Rome, 00184, Italy}
\affiliation{$^{6}$INO-CNR Pitaevskii BEC Center and Dipartimento di Fisica, Universit\`a di Trento, 38123 Povo, Italy}
\affiliation{$^{7}$ Universit\'e Bourgogne Europe, CNRS, Laboratoire Interdisciplinaire Carnot de Bourgogne ICB UMR 6303, 21000 Dijon, France}


\begin{abstract}
Optical thermalization has been recently studied theoretically and experimentally in the 2D spatial evolution of (quasi-)monochromatic light waves propagating in multimode fibers. 
In this work, we investigate the  spatio-temporal equilibrium properties of incoherent multimode  optical waves through the analysis of the (2+1)D Bose-Einstein thermal distribution and the corresponding classical Rayleigh-Jeans approximation. 
In the classical regime, we perform numerical simulations of the nonlinear Schr\"odinger equation and demonstrate relaxation toward the spatio-temporal Rayleigh–Jeans equilibrium state, as described by the corresponding wave turbulence kinetic equation. Remarkable adiabatic cooling phenomena stemming from the high-frequency tails of the Rayleigh-Jeans distribution are discussed and the consequent limitations of the classical approximation are highlighted. To overcome these issues and properly include quantum effects, we make use of a quantum version of the nonlinear Schr\"odinger equation that is obtained from the general quantum theory of light propagating in nonlinear waveguides. The associated kinetic equation describes relaxation toward the spatio-temporal Bose-Einstein equilibrium distribution with a fully regular UV behaviour. The analysis of thermodynamic equilibrium properties reveals a strong dependence on the specific dispersion regime under consideration. In the anomalous dispersion regime, the system relaxes to positive-temperatures equilibrium states: as the number of modes of the waveguide increases, the fundamental spatial mode becomes macroscopically populated, while its temporal spectrum undergoes significant narrowing, ultimately leading to complete (2+1)D spatio-temporal condensation in the thermodynamic limit. By contrast, in the normal dispersion regime the system evolves toward negative-temperature equilibrium states characterized by a hybrid structure: the spatial equilibrium displays an inverted modal population, whereas the temporal spectrum remains peaked around the fundamental (carrier) optical frequency. In this regime, we predict that spatio-temporal light waves exhibit a phase transition to Bose-Einstein condensation at negative temperatures, which occurs by increasing the temperature above a negative critical value. Our work opens new avenues for future research, including the possibility for a dual spatio-temporal beam cleaning through full spatio-temporal light condensation, and lay the groundwork for the development of spatio-temporal optical thermodynamics.
\end{abstract}


\maketitle

\section{Introduction}

There is a growing interest in studying quantum fluids of light with effective photon-photon interactions, which may be viewed as the photonic counterpart to atomic Bose gases \cite{carusotto13,bloch22,glorieux23,glorieux25,wrightNP22,ferraroAPX23}. The concept of fluid of light emerges from a mathematical mapping between the propagation of a light field through a nonlinear medium and the conservative temporal evolution of a quantum fluid of interacting photons~\cite{larre15}.  Such propagating fluids of light have been used to explore a wide range of intriguing phenomena, e.g., the generation of superfluid Bogoliubov sound waves \cite{fontaine18,michel18}, complex vortex dynamics \cite{fleischerBKT,michel21,glorieux23b,congy24,pavloff25}, binary fluids~\cite{Piekarski25}, spin-orbit coupled fluids~\cite{martone21,martone23}, analogue gravity and cosmology \cite{segev15,braidotti22,marcucci19,roger16,marino19,shah23}, turbulence cascades regimes \cite{ferreira22,rasooli23}, or the dynamical formation of prethermalized equilibrium states \cite{berges04,larre16,abuzarli22,glorieux22}.

Among the physical phenomena associated with the propagation of fluids of light, full three-dimensional (3D) quantum thermalization to a Bose-Einstein (BE) distribution and photon condensation have been predicted for bulk Kerr media since 2016 \cite{chiocchetta16}. However, because collisional scattering processes between single photons are exceedingly rare in standard media, these effects were also anticipated to typically require prohibitively long propagation lengths. The situation is much different for classical fields, for which thermalization and condensation are anticipated to be dramatically accelerated 
by bosonic stimulation effects~\cite{davis01,PRL05,blakie08,krstulovic11,nazarenko11,PR14}.


In this context, the thermalization of classical optical waves toward the Rayleigh–Jeans (RJ) equilibrium distribution has attracted considerable attention in recent years. In particular, when restricting the analysis to purely monochromatic optical beams -- thereby neglecting temporal dynamics and focusing exclusively on the transverse spatial degrees of freedom -- the problem reduces to a 2D spatial thermalization process. Within this framework, 2D thermalization toward the classical RJ distribution has been predicted for optical waves propagating in multimode fibers \cite{aschieri11}, on the basis of the wave turbulence theory \cite{zakharov92,nazarenko11,Newell-Rumpf,shrira-nazarenko13,PR14,PRL19}. The finite number of modes of an optical beam in guided configuration accelerates the rate of thermalization to the RJ equilibrium, and also regularizes the ultraviolet (UV) catastrophe inherent to classical waves. It is in this guided-wave configuration that RJ thermalization has recently been experimentally studied in the transverse 2D spatial dynamics of a quasi-monochromatic speckle beam that propagates in a multimode fiber \cite{PRL20,EPL21,pourbeyram22,mangini22,podivilov22,mangini23}. This process is associated with the spatial beam self-cleaning effect \cite{wright16,krupa17}, whose physical mechanism has been investigated in different theoretical works \cite{PRL19,podivilov19,PRA19,christodoulides19,sidelnikov19,parto19,makris20,makris22}, see also the reviews \cite{ferraro23,ferraro25_rev}. 
From a broader perspective, this intense activity has given rise to the emerging topical area of optical thermodynamics \cite{christodoulides19,wright22,kottos20,kottos23,ren23,efremidis24_press,efremidis24,ren25}. In this context, different important achievements have been obtained, among which we may highlight the prediction \cite{christodoulides19} and experimental observation of negative temperature equilibrium states \cite{PRL23,muniz23,ferraro25}, the calorimetry of photon gases \cite{ferraro24}, their Joule-Thomson expansion \cite{kirsch25}, and the development of non-equilibrium thermodynamic approaches \cite{PRL22,lu24,kottos24,liu25,efremidis25}. 

Going beyond these works on monochromatic beams, increasing attention is now being devoted to the complex spatio-temporal (ST) dynamics of optical beams propagating in multimode fibers, see, e.g., \cite{agrawal,wrightPRL15,krupa16,wright16,krupa16b,kibler21,mas-arabi18,leventoux21,gentyNC22,kibler23}. 
The one-dimensional turbulent dynamics of purely temporal incoherent optical waves was originally investigated in single mode fibers within the framework of the wave turbulence theory \cite{babin07,OE09,pitois06,lagrange07,PRL10,OE11,PRA15,turitsyn13,churkin15}. In this context, temporally incoherent sources -- such as amplified spontaneous emission (ASE) -- are commonly employed in optical experiments (e.g., \cite{pitois06,PRL10,OE11,PRA15}). This is in contrast with spatial beam cleaning experiments, where pulsed coherent laser sources are used \cite{ferraro23,ferraro25_rev}, in particular to investigate the role of temporal effects in that phenomenon \cite{mangini22,krupa18,labaz24,genty25}. 

Our general objective  is to lay a groundwork for studying the spatio-temporal (ST) thermodynamic equilibrium properties of guided light waves. 
Within the broad context of optical thermalization, the general form of ST RJ equilibrium distribution \cite{OE07} was derived from first principles in Ref.\cite{podivilov22} for multimode fibers, based on  the conservation laws governing optical wave propagation. In a recent work~\cite{zanaglia25}, the nonequilibrium process of ST thermalization in a multimode waveguide was investigated, showing that spatial discrete modes and continuous temporal frequencies cooperate in a joint manner, leading to a significant acceleration of light thermalization towards ST equilibrium. 

In the present article, we specifically investigate the ST equilibrium properties of incoherent multimode optical waves, building on recent advances in the study of purely spatial 2D equilibrium systems~\cite{zanaglia24}. By extending this framework to the (2+1)D spatio-temporal setting, we establish a unified description of equilibrium dynamics in multimode guided waves and reveal qualitatively distinct behaviors governed by different dispersion regimes.

We start by investigating the classical regime within the framework of the ST nonlinear Schr\"odinger (NLS) equation. Numerical simulations reveal thermalization toward the ST RJ equilibrium state, in agreement with the predictions of the associated kinetic equation. Peculiar phenomena stemming from the UV behavior of the classical RJ distribution are illustrated and the limitations of the classical theory are discussed. To overcome these, we then start from the general quantum theory of light propagation in nonlinear media developed in~\cite{larre15} to derive a quantum version of the NLS formalism. In this way, we show that in the quantum regime the corresponding kinetic equation naturally describes relaxation toward the ST BE equilibrium distribution.

The analysis of the equilibrium distributions reveals different properties depending on the nature of the dispersion regime. In the anomalous dispersion regime we show that, as the waveguide becomes highly multimode, the fundamental mode becomes macroscopically populated while its spectrum undergoes significant spectral narrowing, ultimately leading to a complete (2+1)D spatio-temporal condensation in the thermodynamic limit. On the other hand, in the normal dispersion regime we show that negative temperatures emerge as natural equilibrium states of ST thermalization. These states are characterized by a crossed hybrid nature: in the spatial domain the distribution is inverted, with high-order modes more populated than low-order modes, whereas in the time domain the low-frequency components (relative to the carrier frequency) are more populated than the high-frequency components. By increasing the optical temperature above a negative critical value, we show that the ST equilibrium distribution undergoes a phase transition to BE condensation at negative temperatures.

This article is structured as follows. In Sec.~\ref{sec:ST_eq}, we introduce the ST NLS equation and the corresponding wave turbulence kinetic equation, which describes nonequilibrium thermalization toward the ST RJ equilibrium distribution. We corroborate this relaxation dynamics through numerical simulations of the NLS equation, for both positive and negative temperature regimes. In Sec.~\ref{sec:quantum}, we extend the analysis to the quantum regime by considering a quantum version of the NLS formalism, whose associated kinetic equation describes thermalization to the ST BE equilibrium distribution. In the subsequent sections, we investigate the thermodynamic equilibrium properties.
Section~\ref{sec:pos_temp} focuses on the case of positive temperatures through the analysis of the transition to light condensation in the spatial and ST cases. In doing so, we show that the ST configuration provides a natural distinction between the classical RJ regime and the quantum BE regime. The analysis is extended to negative temperatures in Sec.~\ref{sec:BEC_Tneg_grin}, which is essentially devoted to examining the phase-transition to BE condensation at negative temperatures. Finally, the concluding section discusses possible perspectives and future directions for this work.

\begin{center}
\begin{figure}[h]
\includegraphics[width=1\columnwidth]{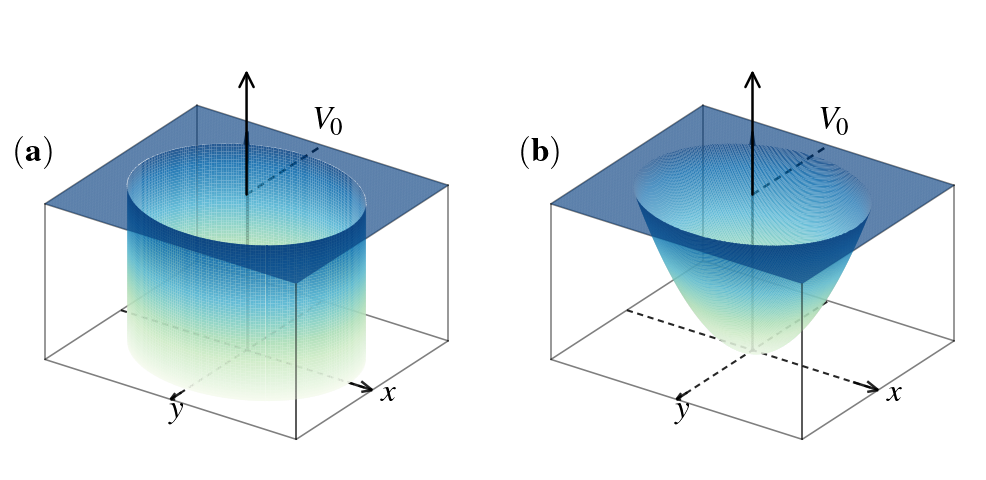}
\caption{
\baselineskip 10pt
{\bf Truncated trapping potentials.} Schematic illustration of the transverse potential $V(x,y)$ in the NLS Eq.(\ref{eq:nls_rt}), which confines the optical field in the plane transverse to the propagation direction $z$. In the waveguide configurations considered here, the potential is truncated at $V_0$: (a) step-index waveguide; (b) parabolic (graded-index) waveguide.
}
\label{fig:potentials}
\end{figure}
\end{center}

\section{Classical nonlinear Schr\"odinger model: thermalization to a spatio-temporal Rayleigh-Jeans distribution}
\label{sec:ST_eq}

We start by considering a classical description of light propagation in a multimode waveguide. This approach is justified when the waveguide modes are highly occupied, a condition that will be made precise below and that allows to formulate the light propagation in terms of a classical field equation for the field amplitude.

\subsection{NLS model}

We consider a continuous spatio-temporally (ST) incoherent optical wave with central (carrier) frequency ${\tilde \omega}_0$, propagating along the $z$-axis of a waveguide, which confines the field in the transverse plane through a trapping potential $V(\br_\perp)$, with $\br_\perp = (x,y)$, see Fig.\ref{fig:potentials}. 
Within a paraxial and narrow-band approximation, we describe light propagation in terms of the NLS equation~\cite{larre15,chiocchetta16,glorieux25}, a classical model that is known to provide a general framework for the spatio-temporal evolution of the slowly varying field envelope $\psi(t,\br_\perp,z)$:
\begin{align}
i\partial_z \psi = -\alpha_\perp \nabla_\perp^2 \psi + V(\br_\perp) \psi + \kappa \partial_t^2 \psi - \gamma  |\psi|^2\psi,
\label{eq:nls_rt}
\end{align}
where the $\nabla_\perp$ operator refers to the transverse $(x,y)$ directions.

In this framework, the propagation distance $z$ then plays the role of an evolution `time' variable, while $t$ plays the role of a spatial variable in the reference frame that propagates with group-velocity $v_g^{-1}=\partial_{{\tilde \omega}} k({\tilde \omega}_0)$, where $k({\tilde \omega})$ is the dispersion relation as a function of the frequency ${\tilde \omega}$, with $k_0=k({\tilde \omega}_0)={\tilde \omega}_0/c_0$, $c_0=c/n_{wg}$ being the light speed with $n_{wg}$ the waveguide  refractive index. The parameter $\kappa=\frac{1}{2} \partial_{\tilde \omega}^2 k({\tilde \omega}_0)$ denotes the dispersion coefficient, $\gamma$ the nonlinear coefficient, and $\alpha_\perp = 1/(2k_0)$ the diffraction coefficient.

In the following, we consider an incoherent guided beam that exhibits fluctuations that are statistically stationary along the longitudinal temporal axis.
Denoting by $\left< \cdot \right>$ an average over the realizations of the random initial conditions $\psi(t,\br_\perp,z=0)$,
the power $P$ defined by
\begin{equation}
{P}= \int d\br_\perp \left< |\psi|^2\right> \, ,
\end{equation} 
 is a conserved quantity: it is independent of $t$ by stationarity and independent of $z$ by the NLS Eq.(\ref{eq:nls_rt}).
Similarly the Hamiltonian ${H}={E}+{U}$ is conserved, which has a linear contribution 
\begin{equation}
{E}=  \int d\br_\perp  \alpha_\perp  \left< |\nabla_\perp \psi|^2\right> -\kappa \left< | \partial_t\psi|^2 \right>+  V(\br_\perp)\left< |\psi|^2 \right> \, ,
\end{equation}
and a nonlinear contribution 
\begin{equation}    
{U} = -\frac{\gamma}{2} \int d\br_\perp \left< |\psi|^4 \right>\,.
\end{equation}
If we consider a time window $T_0$ that is much larger than the correlation time of light $\tau_c$, then
time averaged quantities are equivalent to statistical averages, so that we also have 
\begin{equation}
\frac{1}{T_0}\int_0^{T_0} dt \int d\br_\perp |\psi|^2  \simeq P\, ,
\end{equation} 
when $T_0 \gg \tau_c$,
and similarly for $E$ and $U$.

By introducing the real-valued orthonormal basis $u_m(\br_\perp)$, with eigenvalues $\beta_m$, which are solutions of 
\begin{align}
\beta_m u_m(\br_\perp)=-\alpha_\perp \nabla_\perp^2 u_m(\br_\perp) + V(\br_\perp) u_m(\br_\perp),
\label{eq:eigen}
\end{align} 
and performing a Fourier transform to frequency-space,
we can expand the field on the waveguide modes
\begin{equation}
\label{eq:Four}
    \psi(t,\br_\perp,z)= \frac{1}{2\pi} \sum_m  u_m(\br_\perp) \int b_m({\omega},z) \exp(-i {\omega}t) d{\omega}
\end{equation}
where the frequency-space modal amplitudes are given by 
\begin{equation}
    b_m({\omega},z)  = \iint \psi(t,\br_\perp,z) \exp( i {\omega} t) dt \, u_m(\br_\perp) d\br_\perp,
\end{equation}
and $\omega= {\tilde \omega}-{\tilde \omega}_0$
is the frequency offset relative to the carrier frequency ${\tilde \omega}_0$.
We refer the reader to the note \cite{beta_m_note} for relevant examples of the eigenvalues $\{\beta_m\}$ involving a homogeneous step-index waveguide and a parabolic waveguide. 

The modal amplitudes $b_m({\omega},z)$ satisfy the modal NLS equation:
\begin{align}
i \partial_z {b}_m = {\tilde \beta}_m({\omega})   {b}_m  -  \gamma  Q_{m}({\bm b}),
\label{eq:nls}
\end{align}
with the nonlinear term
\begin{align}
Q_{m}({\bm b}) =& \frac{1}{(2\pi)^2}\sum_{pqr} W_{mpqr} \int {b}_p({\omega}_1,z) {b}_q^*({\omega}_2,z){b}_r({\omega}_3,z)
\nonumber \\ 
&\times \delta({\omega}-{\omega}_1+{\omega}_2-{\omega}_3) d{\omega}_1 d{\omega}_2 d{\omega}_3,
\label{eq:nls_nl}
\end{align}
and the spatio-temporal modal dispersion relation 
\begin{align}
{\tilde \beta}_m({\omega})=\beta_m - \kappa {\omega}^2.
\label{eq:beta_tilde}
\end{align}
The tensor in Eq.(\ref{eq:nls_nl}) 
\begin{equation}
    W_{mpqr} = \int u_m(\br_\perp)u_p(\br_\perp)u_q(\br_\perp)u_r(\br_\perp) d\br_\perp
\end{equation}takes into account for the spatial overlap among the eigenmodes. In a multimode waveguide, the potential is truncated $V(\br_\perp) \le V_0$ as depited in Fig.\ref{fig:potentials}, and guides a finite number $M$ of modes, labeled by the indices $m,p,q,r=0,..,M-1,$, where $M$ is the total number of propagative modes supported by the waveguide.
The parameter $V_0={\rm max}(\beta_m)$ then denotes the depth of the waveguide trapping potential in the transverse dimension.

\subsection{Spatio-temporal RJ equilibrium distribution \label{sec:waveturb}}

The irreversible process of thermalization to equilibrium can be described in detail by the wave turbulence theory in the weakly nonlinear regime $|{U}/{E}| \ll 1$, which is the regime studied in the experiments of optical thermalization  \cite{PRL20,EPL21,pourbeyram22,mangini22,podivilov22,mangini23}.

In a recent work \cite{zanaglia25}, a ST kinetic equation has been derived from the NLS Eq.(\ref{eq:nls}) by using the wave turbulence theory. 
For the sake of completeness and to facilitate comparison with the corresponding quantum version discussed in Sec.~III, we report here the classical kinetic equation derived in Ref.\cite{zanaglia25}.
The averaged modal components ${\tilde n}_m(\omega, z)$ are defined by 
\begin{equation}
    \left< {b}_m(\omega, z) {b}_{p}^*(\omega', z) \right> ={\tilde n}_m(\omega, z) \delta_{mp} \delta(\omega-\omega')\,,
\end{equation} 
where $\delta_{mp}$ is the Kronecker symbol.
The kinetic equation describes their nonequilibrium evolution in $z$:
\begin{align}
\partial_z {\tilde n}_m(\omega,z)=& 4\pi \gamma^2 \sum_{pqr} \int d\omega_{1-3} |W_{mpqr}|^2 {\tilde {\bm M}}_{mpqr}({\bm {\tilde n}}) 
\nonumber \\ 
&\times \delta(\omega-\omega_1+\omega_2-\omega_3) \delta(\Delta {\tilde \beta}_{mpqr}^{\omega 123}) ,
\label{eq:kin_class}
\end{align}
where 
\begin{align*}
{\tilde {\bm M}}_{mpqr}({\bm {\tilde n}})=& \big( {\tilde n}_m(\omega)+ {\tilde n}_q(\omega_2) \big) {\tilde n}_p(\omega_1) {\tilde n}_r(\omega_3) 
\nonumber \\
&- {\tilde n}_m(\omega) {\tilde n}_q(\omega_2)  \big( {\tilde n}_p(\omega_1)+ {\tilde n}_r(\omega_3) \big),
\end{align*}
and 
\begin{align*}
\Delta {\tilde \beta}_{mpqr}^{\omega 123}={\tilde \beta}_m(\omega)-{\tilde \beta}_p(\omega_1)+{\tilde \beta}_q(\omega_2)-{\tilde \beta}_r(\omega_3)
\end{align*}
is the four mode spatio-temporal frequency resonance.
The kinetic Eq.(\ref{eq:kin_class}) conserves the power
\begin{equation}
P = \sum_m  \frac{1}{(2\pi)^2} \int {\tilde n}_m(\omega,z) d\omega,
\end{equation}
and the linear contribution to the energy  
\begin{equation}
E = \sum_m  \frac{1}{(2\pi)^2}\int {\tilde \beta}_m(\omega) {\tilde n}_m(\omega,z) d\omega.    
\end{equation}
Note that $E \simeq H$ in the considered weakly nonlinear regime $|U/E| \ll 1$. Also note that the factor $1/(2\pi)^2$ comes from the adopted Fourier transform convention, see Eq.(\ref{eq:Four}).

Furthermore, the ST kinetic equation is formally irreversible in the evolution (`time') variable $z$, as expressed by an $H-$theorem of entropy growth $\partial_z {\tilde {\cal S}}(z) \ge 0$, where the nonequilibrium entropy is 
\begin{equation}
{\tilde {\cal S}}(z)= \sum_m  \frac{1}{(2\pi)^2} \int \log\big({\tilde n}_m(\omega,z)\big) d\omega.
\end{equation} 
Then in contrast to the NLS Eq.(\ref{eq:nls}) that is formally reversible (Hamiltonian), the ST kinetic equation describes the actual irreversible process of thermalization toward the RJ equilibrium.

The ST RJ distribution is obtaied by maximizing  the entropy ${\tilde {\cal S}}$, under the constraints imposed by the conservation of the energy ${E}$ and the power ${P}$, leading to
\begin{equation}
{\tilde n}_m^{RJ}(\omega) = \frac{\tilde T}{{\tilde \beta}_m(\omega)-{\tilde \mu}} ,
\label{eq:n_rj_general}
\end{equation}
where ${\tilde T}$ and ${\tilde \mu}$ are the temperature and the chemical potential, i.e., the Lagrange multipliers associated to the conservation of $(E, P)$.
We recall here that we are dealing with a closed Hamiltonian system (i.e., microcanonical statistical ensemble without a thermostat), so that ${\tilde T}$ and ${\tilde \mu}$ are not externally imposed parameters but are fully determined by the conserved quantities $(E,P)$ \cite{zanaglia25}.

The physical condition ${\tilde n}_m^{RJ}(\omega) >0$ in Eq.(\ref{eq:n_rj_general}), yields the following conditions on the thermodynamic parameters. If $\kappa < 0$ (anomalous dispersion), temporal dispersion and spatial diffraction operate in the same way, and the equilibrium is characterized by a positive temperature,  ${\tilde T}>0$ with ${\tilde \mu} < \beta_0$. Conversely, if $\kappa > 0$ (normal dispersion), temporal dispersion and spatial diffraction operate in opposite ways, and the equilibrium is characterized by a negative temperature ${\tilde T}<0$ with ${\tilde \mu} > V_0$, where we recall that  $V_0={\rm max}(\beta_m)$ denotes the depth of the waveguide trapping potential. This simple observation shows that negative temperature equilibria emerge as natural equilibrium states of ST thermalization in the normal dispersion regime. Note that this conclusion is not modified by accounting for a group-velocity difference among the modal components, see Appendix~\ref{app:B}.

To summarize, we have:
\begin{align}
&\kappa < 0 \quad \to \quad  {\tilde T}> 0, \ {\tilde \mu} < \beta_0,
\label{eq:sign_kappa1}
\\
&\kappa > 0 \quad \to \quad   {\tilde T}< 0, \ {\tilde \mu} > V_0.
\label{eq:sign_kappa2}
\end{align}
We anticipate that negative temperature ST equilibria are characterized by a crossed hybrid nature: while in the spatial domain the population is inverted (the highest waveguide mode $\beta_{\rm max}$ being the most populated), in the temporal domain the fundamental frequency component ($\omega=0$) is the most populated.

\subsection{NLS simulations}

We confirm the equilibrium properties (\ref{eq:sign_kappa1}-\ref{eq:sign_kappa2}) at positive and negative dispersion regimes by direct numerical simulations of the NLS equation with periodic boundary conditions in time over a time window $T_0$ much larger than the correlation time of light, $\tau_c \ll T_0$. For this purpose, we have considered step-index waveguides with different characteristics (e.g., waveguide radius, or depth of trapping potential $V_0$, see Fig.~\ref{fig:potentials}), since the normal and anomalous dispersion regimes typically occur at different wavelengths, which in turn modify the waveguide characteristics and corresponding propagation constants. 
For instance, this is the case for silica fibers, where the zero-dispersion wavelength is around $\lambda_0 \simeq 1.3 \mu m$ -- the dispersion regime being anomalous (normal) for $\lambda > \lambda_0$ ($\lambda < \lambda_0$).
It is worth noting that, beyond multimode optical fibers, different alternative waveguide configurations and experimental platforms are available, including, e.g., optically induced waveguides in photorefractive crystals \cite{denz03} and atomic vapors \cite{glorieux22}.
In photorefractive SBN crystals, the zero dispersion wavelength is also around $\lambda_0 \simeq 1.3 \mu m$  --  dispersion being normal (anomalous) for $\lambda < \lambda_0$ ($\lambda > \lambda_0$) \cite{abarkan08}.
In atomic vapors, the dispersion can be tuned from positive to negative in the vicinity of an atomic resonance \cite{Abuzarli_PhD}, and its magnitude can greatly exceed that encountered in standard optical fibers \cite{glorieux25}.

\begin{center}
\begin{figure}[h]
\includegraphics[width=1\columnwidth]{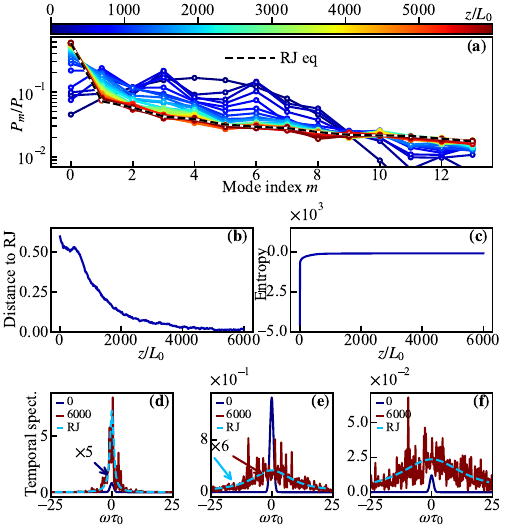}
\caption{
\baselineskip 10pt
{\bf Spatio-temporal thermalization to positive temperatures.}
Simulation of the NLS Eq.(\ref{eq:nls}) in the anomalous dispersion regime, $\kappa < 0$. 
(a) Evolution of spatial modal occupation $P_m(z)/P$, showing the relaxation to the equilibrium RJ distribution Eq.(\ref{eq:n_rj_general}) (dashed black line).
(b) Evolution of the distance ${\cal D}(z)$ to equilibrium, whose decrease to zero evidences attraction to the RJ equilibrium. 
(c) This irreversible process is  characterized by a monotonous growth of entropy, as described by the $H-$theorem of the wave turbulence kinetic Eq.(\ref{eq:kin_class}).
Temporal spectrum $|b_m(\omega, z)|^2$  of the fundamental mode $m=0$ (d), intermediate mode $m=7$ (e), highest mode ($m=13$) (f), at $z=0$ (dark blue) and $z = 6000 L_{0}$ (red), showing thermalization to RJ spectra (dashed light blue). Parameters are given in the text.
}
\label{fig:nls_anom}
\end{figure}
\end{center}

\begin{center}
\begin{figure}[h]
\includegraphics[width=1\columnwidth]{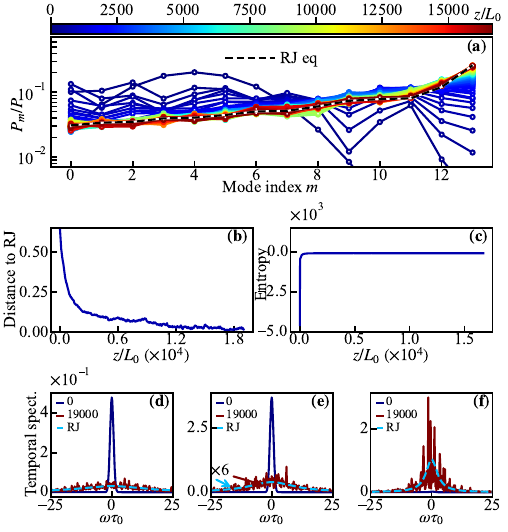}
\caption{
\baselineskip 10pt
{\bf Spatio-temporal thermalization to negative temperatures.}
Simulation of the NLS Eq.(\ref{eq:nls}) in the normal dispersion regime, $\kappa > 0$. 
(a) Evolution of spatial modal occupation $P_m(z)/P$, showing the relaxation to the equilibrium RJ distribution Eq.(\ref{eq:n_rj_general}) (dashed black line).
Note that, in contrast to positive temperatures (Fig.~\ref{fig:nls_anom}(a)), at negative temperatures the equilibrium spatial spectrum is inverted, with high-order modes more populated than low-order modes.
(b) Evolution of the distance ${\cal D}(z)$ to equilibrium, whose decrease to zero evidences attraction to the RJ equilibrium. 
(c) This irreversible process is  characterized by a monotonous growth of entropy, as described by the $H-$theorem of the wave turbulence kinetic Eq.(\ref{eq:kin_class}).
Temporal spectrum $|b_m(\omega, z)|^2$  of the fundamental mode $m=0$ (d), intermediate mode $m=7$ (e), highest mode ($m=13$) (f), at $z=0$ (dark blue) and $z = 19 \, 000 L_{0}$ (red), showing thermalization to RJ spectra (dashed light blue). Parameters are given in the text.
}
\label{fig:nls_norm}
\end{figure}
\end{center}

We have performed simulations of the NLS Eq.(\ref{eq:nls}) in the anomalous and the normal dispersion regimes, 
whose results are reported in Fig.~\ref{fig:nls_anom} and Fig.~\ref{fig:nls_norm}, respectively.
We considered step-index waveguides supporting $M=14$ modes at $\lambda=0.5 \mu$m in the normal dispersion regime ($V_0=5654.00$m$^{-1}$ and cross-sectional waveguide surface $S=8.0\times10^{-10}$m$^2$), and $\lambda=2.4 \mu$m in the anomalous dispersion regime ($V_0=1177.91$m$^{-1}, S=2.0\times 10^{-8}$m$^2$).
The initial condition is a spatio-temporally incoherent field, in which the different modal components $b_m(\omega, z=0)$ are independent complex-valued Gaussian random variables of zero mean; each spatial mode $m$ has a Gaussian spectrum $\sim \exp(-\omega^2/\sigma_\omega^2)$, with the same width $\sigma_\omega = 1/\tau_0$ for all modes, and different amplitudes, see panels (a) in Figs.~\ref{fig:nls_anom}-\ref{fig:nls_norm}. The parameter $\tau_0=\sqrt{|\kappa| L_0}$ denotes the `healing time', where $L_{0}=1/(|\gamma| I_0)$ is the nonlinear length when all the power is confined in the fundamental mode, with intensity $I_0=P/S_{0}$, $S_{0} = 1/W_{0000}$ being the effective area of the fundamental mode. The eigenvalues $\{\beta_m\}$ are computed by solving numerically Eq.(\ref{eq:eigen}). We consider the weakly nonlinear regime $|U/E| \ll 1$ so as to avoid the formation of strongly nonlinear coherent structures, which can freeze the thermalization process \cite{nazarenko11,PR14} (also see the discussion in the conclusion). Specifically, we consider a weakly nonlinear spatial mode dynamics $\beta_m L_{0} \gg 1$ with $L_{0}=0.3$m in the anomalous dispersion regime and $L_{0}=0.05$m in the 
normal dispersion regime  (the eigenvalues lie within the interval $\beta_0 = 28.70/L_{0}$ and $\beta_{M-1} \simeq 338.81/L_{0}$ for $\kappa < 0$, and between $\beta_0 = 23.90/L_{0}$ and $\beta_{M-1} \simeq 277.61/L_{0}$ for $\kappa >0$). We also consider a weakly nonlinear temporal dynamics, which is characterized by a correlation time $\tau_c$ of the field  typically much smaller than the `healing time' $\tau_0$. In particular, in this regime, the formation of temporal solitons is expected to be suppressed. This is important because the presence of solitons is known to slow, or even arrest, the thermalization process. To further ensure the absence of soliton generation, we consider a defocusing nonlinearity ($\gamma < 0$) in the anomalous dispersion regime, and a focusing nonlinearity ($\gamma > 0$) in the normal dispersion regime, so that $\gamma \kappa >0$ in both Fig.~\ref{fig:nls_anom} and Fig.~\ref{fig:nls_norm}. 
We refer the reader to the Supplementary Material of Ref.\cite{zanaglia25} for further details about the method employed to numerically integrate the modal NLS Eq.(\ref{eq:nls}).

Thermalization toward the ST RJ equilibrium [Eq.~(\ref{eq:n_rj_general})] is illustrated in Figs.~\ref{fig:nls_anom}–\ref{fig:nls_norm}. According to the analysis discussed through (\ref{eq:sign_kappa1}–\ref{eq:sign_kappa2}), the system relaxes toward a positive-temperature equilibrium in the anomalous-dispersion regime (Fig.~\ref{fig:nls_anom}), whereas it relaxes to a negative-temperature equilibrium in the normal-dispersion regime (Fig.~\ref{fig:nls_norm}).
The thermalization process is first characterized in the spatial domain through the modal population
${P}_m(z)=\frac{1}{(2\pi)^2}\int {\tilde n}_m(\omega,z) d\omega$, obtained by integrating over temporal frequencies, with $P=\sum_m P_m(z)$ the conserved power. A key distinction between the normal- and anomalous-dispersion regimes in Figs.~\ref{fig:nls_anom}–\ref{fig:nls_norm} is the inversion of the spatial modal population, which is peaked on the highest spatial mode at negative temperatures, as predicted above through (\ref{eq:sign_kappa1}-\ref{eq:sign_kappa2}). Thermalization is also evidenced in the temporal domain by the convergence of the spectra toward the RJ prediction, as illustrated in panels (d)-(e)-(f) of Figs.~\ref{fig:nls_anom}–\ref{fig:nls_norm} for the spatial modes $m=0, 7, 13$. A quantitative measure of thermalization is provided by the distance to equilibrium
\begin{equation}
{\cal D}(z)=\frac{\sum_m \left|{P}_m(z)-{P}_m^{RJ}\right|}{\sum_m \left({P}_m(z)+{P}_m^{RJ}\right)},
\end{equation}
where ${P}_m^{RJ}=\frac{1}{(2\pi)^2}\int {\tilde n}_m^{RJ}(\omega) d\omega$ denotes the  spatial RJ modal distribution.
By construction, ${\cal D}$ is bounded as $0 \le {\cal D} \le 1$. As shown in panels (b) of Figs.~\ref{fig:nls_anom}-\ref{fig:nls_norm}, ${\cal D}(z)$ decreases toward zero along propagation, indicating convergence to equilibrium.
Thermalization is further corroborated by the growth of entropy, which saturates as the field approaches the RJ state, see panels (c) in Figs.~\ref{fig:nls_anom}-\ref{fig:nls_norm}.

The parameters $({\tilde T},{\tilde \mu})$ in Eq.~(\ref{eq:n_rj_general}) are evaluated by accounting for the spectral cutoff \textcolor{blue}{$\omega_c \tau_0 = 25$} imposed by the numerical grid (see Ref.~\cite{zanaglia25} for details).
Thus, the finite spectral cutoff arising from the numerical discretization of the temporal grid regularizes the UV divergence inherent to classical waves, as discussed further in the next section.


\begin{center}
\begin{figure}[!t]
\includegraphics[width=.9\columnwidth]{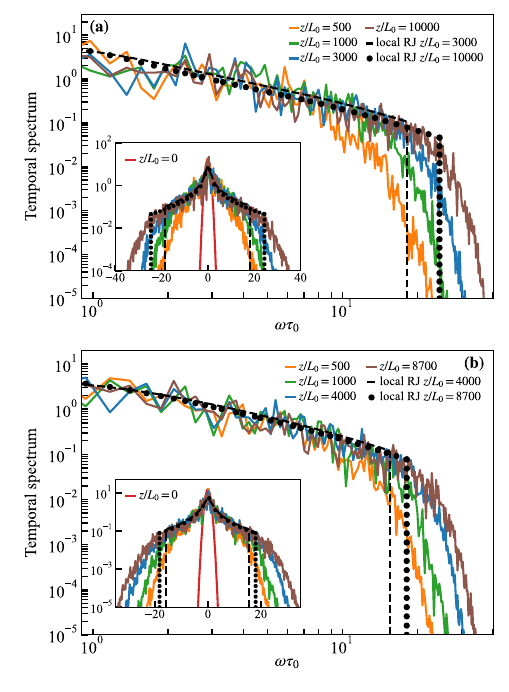}
\caption{{\bf Spectral broadening following a quasi-equilibrium process of ST thermalization in anomalous and normal dispersion regimes.}
Temporal spectra (summed over the individual modal spectra, $\sum_m |b_m(\omega,z)|^2$) obtained from simulations of the NLS Eq.(\ref{eq:nls}) in the anomalous dispersion regime $\kappa < 0$ (a), and in the normal dispersion regime $\kappa > 0$ (b), for several propagation distances. The simulations were performed considering the first 5 spatial modes of the respective 14-mode fiber from Fig.~\ref{fig:nls_anom} and Fig.~\ref{fig:nls_norm}. The two insets display the same spectra in semi-logarithmic representation, with the initial spectrum at $z=0$ (red line).
After an initial transient, spectral broadening is characterized by a front that propagates in frequency space, leaving a local RJ distribution in its wake, i.e., the spectra are well described by a truncated RJ distribution (dotted and dashed dark lines), in agreement with the quasi-equilibrium process of ST thermalization reported in Ref.\cite{zanaglia25}.
Note that despite the similar behavior between the two dispersion regimes in (a) and (b), normal dispersion leads to equilibria with negative temperatures ${\tilde T} < 0$, as discussed through (\ref{eq:sign_kappa1}-\ref{eq:sign_kappa2}) and the Figs.~\ref{fig:nls_anom}-\ref{fig:nls_norm}. The parameters are specified in the text.
\baselineskip 10pt}
\label{fig:nls_front}
\end{figure}
\end{center}

\subsection{Local-equilibrium route to ST RJ thermalization and UV catastrophe}
\label{subsec:uv_cat}

A well-known limitation of classical field theories is the emergence of UV divergences in a thermal equilibrium state, commonly referred to as the black-body catastrophe~\cite{zakharov92,nazarenko11,blakie08}.
For our specific problem of ST thermalization in a guided configuration, this issue has been discussed in the recent work~\cite{zanaglia25}, where a quasi-equilibrium process of ST thermalization and adiabatic cooling has been identified. 

In our setup, UV divergences are naturally regularized in the transverse direction by the finite number of guided modes, whereas in the temporal (frequency) domain they give rise to nontrivial physical behaviors.
In the numerical simulations, we consider an initial condition that is strongly nonthermal, characterized by a short-tailed Gaussian spectral distribution. Following an initial transient, it was shown in Ref.\cite{zanaglia25} that the optical field evolves toward a {\it local quasi-equilibrium state} at each propagation distance $z$. This state is accurately described by a truncated RJ equilibrium distribution, established within a finite frequency window $[-\omega_c^{loc}(z), \omega_c^{loc}(z)]$. Outside this window, the spectral density rapidly decays to zero.


During the propagation in $z$, the bounded window $[-{\omega}_{c}^{loc}(z),{\omega}_{c}^{loc}(z)]$ of the local RJ equilibrium expands, and the corresponding local thermodynamic parameters also display a $z$ dependence. In particular, the local temperature ${\tilde T}^{loc}(z)$ was shown to undergo an adiabatic decrease due to the continuous transfer of incoherent beam fluctuations into the high-energy tails of the spectral distribution at constant energy ($E=$const). This mechanism motivates the term adiabatic cooling used to describe the process in Ref.\cite{zanaglia25}. Making use of the wave turbulence kinetic Eq.(\ref{eq:kin_class}), the following scaling law was obtained for the spectral expansion and corresponding temperature decay: $\omega_c^{loc}(z) \sim 1/{\tilde T}^{loc}(z) \sim z^{1/7}$, a theoretical prediction that has been confirmed by NLS simulations \cite{zanaglia25}.

Provided the numerical frequency cut-off $\omega_c$ imposed by the spectral grid is chosen sufficiently large to accommodate the truncated local RJ distribution, i.e., $\omega_{c}^{loc}(z) < \omega_c$, the optical-field spectrum is free to evolve without numerical constraints. 
This is illustrated in Fig.~\ref{fig:nls_front} that shows the evolution during propagation of the mode-integrated temporal spectrum of the field, $\sum_m |b_m(\omega,z)|^2$, in both the anomalous dispersion ($\kappa < 0$, Fig.~\ref{fig:nls_front}(a)), and the normal dispersion ($\kappa >0$, Fig.~\ref{fig:nls_front}(b)). 
In the simulation presented in Fig.~\ref{fig:nls_front}, the number of modes has been reduced to 5, so as to track the evolution over longer propagation lengths within a larger frequency window ($\omega_c = 50/\tau_0$ in Fig.~\ref{fig:nls_front}, instead of $\omega_c = 25/\tau_0$ in Figs.~\ref{fig:nls_anom}-\ref{fig:nls_norm}).
Accordingly, the spectrum progressively broadens as propagation proceeds, and is locally well described by a truncated RJ distribution (dashed and dotted black lines), in agreement with the quasi-equilibrium process of adiabatic cooling described in Ref.~\cite{zanaglia25}. Of course, the spectral distribution stops broadening when the RJ window $\omega_c^{loc}(z)$ hits the numerical cut-off $\omega_c$. At this point, the distribution stabilizes into a complete RJ distribution covering the whole numerical spectral grid.

While this is a numerical feature, this classical thermalization and cooling dynamics has an intrinsic limitation imposed by the range of validity of the classical field model employed here. This classical model in fact fails when the mode occupation drops below unity and the discreteness of the field quanta starts to matter. At this very long propagation length, a full quantum description of the field is then needed, whose equilibrium state has the form of a BE distribution. This automatically cuts the UV tails of the distribution and can be seen as an effective high-frequency cut-off, approximately at the frequency for which the mode occupation predicted by the RJ distribution drops below unity~\cite{davis01,blakie08}. As soon as the edge $\omega_c^{loc}(z)$ of the truncated RJ distribution hits this physical cut-off, the adiabatic cooling stops and light stabilizes into a true thermal equilibrium state.

A theoretical description of thermalization towards a BE distribution and of the equilibrium state was first proposed in~\cite{chiocchetta16} for an untrapped field propagating in a spatially homogeneous, bulk medium. A detailed study of its consequences for our specific multimode waveguide geometry will be the subject of the next Section.

\newpage 

\section{Quantum theory: thermalization to a spatio-temporal Bose-Einstein distribution}
\label{sec:quantum}

\subsection{Quantum Hamiltonian in the $t-z$ mapping}

A fully quantum description of our light propagation problem can be obtained by applying the general quantization formalism developed in~\cite{larre15} and based on the so-called $t-z$ mapping to a waveguide geometry. In this formalism, light propagation along $z$ is described in terms of a many-body Hamiltonian for a quantum Bose field living in a three-dimensional space spanned by the two-dimensional transverse coordinates $\br_\perp$ and the (renormalized) physical time $\zeta=v_g\,t$, and evolving during the temporal variable $\tau=z/v_g$ proportional to the propagation direction $z$.

In formulas, the evolution along $\tau=z/v_g$ is set by a Schr\"odinger-like equation
\begin{equation}
i\hbar\frac{d}{d\tau}|\psi\rangle=\hat{H}|\psi\rangle
\end{equation}
for the many-body wavefunction $|\psi\rangle$ living in the many-body Fock space, where the Hamiltonian $\hat{H}$ has the form of an interacting Bose field
\begin{multline}
{\hat H}=
\frac{1}{\tilde \omega_0}
\int\!d^3R \,\left[\alpha_\perp  \nabla_\perp\psihd\,\nabla_\perp\psih-\kappa  v_g^2\frac{\partial\psihd}{\partial \zeta}\frac{\partial\psih}{\partial \zeta}  \right.+ \\ + \left.V\psihd\,\psih-\frac{\gamma}{2} \psihd\,\psihd\,\psih\,\psih\right]\,.
\label{eq:Hprop}
\end{multline}
in the three-dimensional space ${\bf R}=(x,y,\zeta)$ and the field commutation rules inherit the exchanged roles of the spatial propagation direction $z$ and time $t$,
\begin{equation}
[\psih(\br_\perp,\zeta;\tau),\psihd(\br'_\perp,\zeta';\tau)]=
\hbar{\tilde \omega}_0\,v_g
\,\delta^{(2)}(\br_\perp-\br'_\perp)\,
\delta(\zeta-\zeta')\,.
\end{equation}
It immediately follows that the Heisenberg equation for the quantum field $\hat{\psi}$ recovers a quantum version of the propagation equation \eqref{eq:nls_rt}, namely a quantum nonlinear Schr\"odinger equation,
\begin{multline}
 i \frac{d\psih}{d\tau}=\frac{1}{\hbar}[\psih,\hat{H}]=v_g \left[-\alpha_\perp \nabla_{\perp}^2\psih + \kappa v_g^2\frac{\partial^2 \psih}{\partial \zeta^2} + \right. \\ + \left. V\,\psih- \gamma \,\psihd\psih\psih\right]\,.
 \label{eq:qNLSE}
\end{multline}
where the $v_g$ factors on the right-hand-side account for the $z=v_g \tau$ and $t=\zeta/v_g$ relations of the  $t\leftrightarrow z$ mapping.

\newpage 

\subsection{Relaxation to a quantum ST BE distribution}

This formalism based on the quantum nonlinear Schr\"odinger equations is  naturally able to encode in an {\it ab initio} way the discreteness of the photons without any heuristic assumption and without any restriction to finite-duration pulses. However, an exact solution of the many-body problem (\ref{eq:Hprop}-\ref{eq:qNLSE}) in the general case is beyond the reach of theoretical tools.

For the systems under consideration here, the interactions are weak enough for a quantum kinetic approach to be viable~\cite{zakharov92,kadanoff,pomeau_tran19}.
As in the classical wave turbulence theory of Sec.\ref{sec:waveturb}, the evolution is described here in term of a kinetic equation for the mode occupations, with the difference that the classical occupation number is replaced by the quantum expectation value of the photon number operator.

From the quantum formalism, we can define a photon-number distribution $n_m$ in the mode $m$ per unit effective length $\zeta=v_g t$ at effective conjugate momentum $k_\zeta=\omega/v_g$. Straightforward manipulations summarized in Appendix~\ref{app:A} show that $n_m$ is related to the classical wave distribution ${\tilde n}_m(\omega)$ by
\begin{equation}
n_m(k_\zeta) = \tilde n_m(\omega)/(2\pi \hbar \tilde{\omega}_0)
\label{eq:quant_class_corr_main}
\end{equation}
and that this quantity can be equivalently interpreted as a frequency-space photon-number distribution $n_m(\omega)$ in the mode $m$ per unit time $t$.

Going from the classical kinetic equations to the quantum ones, the collision integrals are modified to account for spontaneous events~\cite{zakharov92,kadanoff,pomeau_tran19}, which introduces the usual bosonic enhancement factor in the scattering terms $n_m \to n_m + 1$. Then starting from the classical kinetic Eq.(\ref{eq:kin_class}), we can write the quantum version of the kinetic equation for $n_m(\omega)$ in the form:
\begin{align}
\partial_z {n}_m(\omega,z)=\Gamma \sum_{pqr} \int d\omega_{1-3} |W_{mpqr}|^2 {\bm M}_{mpqr}({\bm {n}}) 
\nonumber \\ 
\quad \times \delta(\omega-\omega_1+\omega_2-\omega_3) \delta(\Delta {\tilde \beta}_{mpqr}^{\omega 123}) ,
\label{eq:kin}
\end{align}
where $\Gamma = 4\pi \gamma^2 (2\pi\hbar {\tilde \omega}_0)^2$ and 
\begin{align*}
{\bm M}_{mpqr}({\bm {n}})=& \big( {n}_m(\omega)+ {n}_q(\omega_2) + 1\big) {n}_p(\omega_1) {n}_r(\omega_3) 
\nonumber \\
&- {n}_m(\omega) {n}_q(\omega_2)  \big( {n}_p(\omega_1)+ {n}_r(\omega_3) + 1\big).
\end{align*}
The quantum kinetic Eq.(\ref{eq:kin}) conserves the power 
\begin{equation}
P = \hbar {\tilde \omega}_0 \sum_m \int \frac{d\omega}{2\pi} {n}_m(\omega) ,     
\end{equation}
and the energy 
\begin{equation}
E = \hbar {\tilde \omega}_0 \sum_m \int \frac{d\omega}{2\pi} {\tilde \beta}_m(\omega) {n}_m(\omega) .  
\end{equation}
In addition, the kinetic equation exhibits an $H-$theorem of entropy growth $\partial_z {\cal S}(z) \ge 0$ for the nonequilibrium entropy 
\begin{align}
{\cal S}(z)=& \hbar {\tilde \omega}_0 \sum_m  \int  \frac{d\omega}{2\pi} \big[ \big({n}_m(\omega)+ 1\big)  \log\big({n}_m(\omega)+1\big) \nonumber \\
&- {n}_m(\omega)  \log\big({n}_m(\omega)\big) \big] .
\label{eq:kin_ent}
\end{align}
The kinetic Eq.(\ref{eq:kin}) then describes an irreversible process of thermalization to equilibrium.

The equilibrium distribution is obtained by maximizing the entropy under the constraints imposed by the conservation of ${P}$ and ${E}$. Using the method of the Lagrange multipliers, we obtain the ST BE equilibrium distribution:
\begin{equation}
n_m^{BE}(\omega) = \frac{1}{\exp\Big( \frac{{\tilde \beta}_m(\omega) - {\mu}}{{T}}\Big) - 1}.
\label{eq:BE_to_RJ}
\end{equation}
We remark that, as for the classical case discussed above through the RJ distribution (\ref{eq:n_rj_general}), the temperature ${T}$, and the chemical potential ${\mu}$ in (\ref{eq:BE_to_RJ}), are not externally imposed by a thermostat, but are fully determined by the conserved quantities $(E,P)$. 
Note how our formalism and our predictions are distinct from other recent works on BE statistics in the beam-cleaning context: for higher-order modes, the output mode power distribution that accompanies the beam self-cleaning experiments of \cite{mangini22,pourbeyram22} can be fitted by the BE law \cite{zitelli23}, as it was confirmed by  experiments in relatively long spans of graded-index (GRIN) multimode fibers \cite{zitelli23,zitelli24}; furthermore, recent experiments have indicated that beam-cleaning in a dissipative double-clad fiber can be understood in analogy to BE condensation \cite{steinmeyer24}.

In the limit where the modes are highly populated $n_m^{BE}(\omega) \gg 1$, the BE distribution (\ref{eq:BE_to_RJ}) can be approximated by the photon-number RJ distribution 
\begin{equation}
n_m^{RJ}(\omega) = \frac{T}{ {\tilde \beta}_m(\omega) - \mu }.
\label{eq:BE_to_RJ_quant}
\end{equation}
This equilibrium distribution is connected to the RJ equilibrium for classical waves ${\tilde n}_m^{RJ}(\omega)$ [see Eq.(\ref{eq:n_rj_general})] through Eq.(\ref{eq:quant_class_corr_main}), which yields the  relation between the corresponding temperatures $T = {\tilde T} / (2\pi \hbar {\tilde \omega}_0)$. 
On the other hand, the chemical potential in the RJ distribution for classical waves (\ref{eq:n_rj_general}) coincides with that entering the photon-number equilibrium distribution (\ref{eq:BE_to_RJ_quant}), that is $\mu = \tilde\mu$. While the RJ distribution well approximates the low-energy part of the BE distribution for which $n_m(\omega) \gg 1$, it overestimates the high-energy tails of the distribution. The power-law instead of exponential decay of $n_m^{RJ}(\omega)$ at large $\omega$ is responsible for the UV divergence issues mentioned in Sec.\ref{subsec:uv_cat}.

\begin{center}
\begin{figure}[!t]
\includegraphics[width=1\columnwidth]{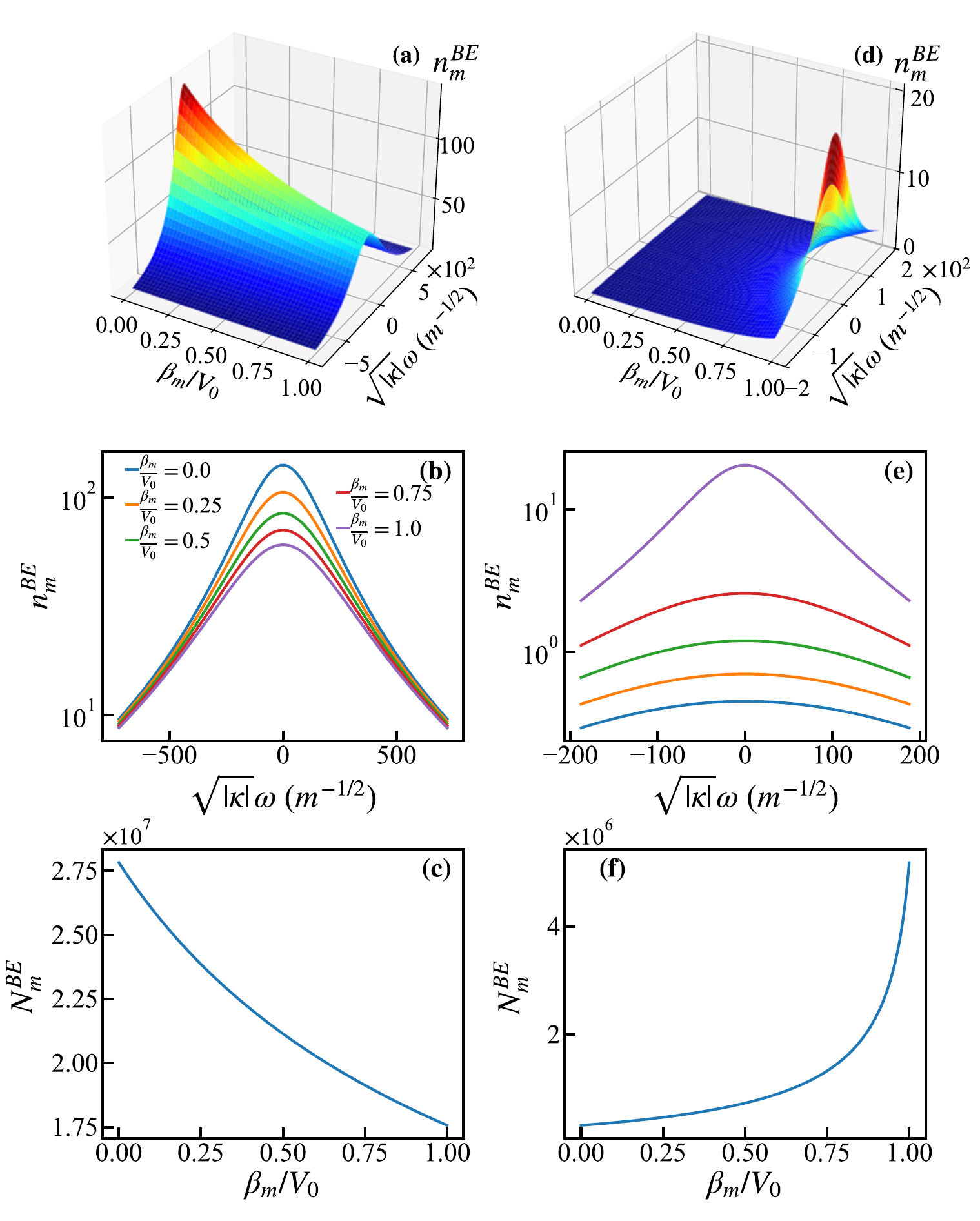}
\caption{
\baselineskip 10pt
{\bf ST equilibria at positive and negative temperatures.} Equilibrium distribution (\ref{eq:BE_to_RJ}) for a step-index multimode waveguide, in the anomalous dispersion regime ($\kappa<0$) leading to positive temperatures, ${T}>0$ (left column), and in the normal dispersion regime  ($\kappa>0$)  leading to negative temperatures, ${T}<0$ (right column). At variance with ${T}>0$ where the spatial and temporal spectra are peaked in the fundamental spatial mode $m=0$ at frequency $\omega=0$, for ${T}<0$ the equilibrium is featured by an inverted population of spatial modes, with temporal spectra still peaked at $\omega=0$.
The equilibrium distributions are plotted at ${T}=1.3{T}_c$ [left column: (a-b-c)], and ${T}=-2|{T}_c|$ [right column: (d-e-f)], where $T_c$ denote the critical temperatures to BE condensation, defined in Eq.(\ref{eq:T_c_pos}) and Eq.(\ref{eq:Tc_neg}) at positive and negative temperatures, respectively.
The plots on the first row report the ST distribution $n_m^{BE}(\omega)$ from Eq.(\ref{eq:BE_to_RJ}), the plots on the second and third rows display cross-sections of the corresponding distributions in the top row: temporal spectra $n_m^{BE}(\omega)$ for different modes with eigenvalues $\beta_m$ (2nd row);  spatial mode distribution integrated in frequency, $N_m^{BE}=(2\pi v_g)^{-1}\int n_m^{BE}(\omega) d\omega$ (3rd row). 
Parameters are given in the text. 
}
\label{fig:no_cond}
\end{figure}
\end{center}

In the remainder of this article, we analyze the equilibrium behavior of the ST BE distribution. Before enetering into details, we note that the distribution (\ref{eq:BE_to_RJ}) satisfies the properties (\ref{eq:sign_kappa1}-\ref{eq:sign_kappa2}) of the RJ equilibrium discussed above through NLS simulations in Fig.~\ref{fig:nls_anom}-\ref{fig:nls_norm}.
We illustrate such properties in Fig.~\ref{fig:no_cond}, for positive and negative temperatures, respectively.
In the examples shown in Fig.~\ref{fig:no_cond}, we have considered for ${T}>0$ (left column) a step-index waveguide with $V_0=52 \,726 \,m^{-1}$ and a waveguide surface $S=1.26 \times 10^{-7} \, m^2$, while for ${T} < 0$ (right column) $V_0=126 \,093 \,m^{-1}$ with $S=3.14\times 10^{-8} \, m^2$.
In both cases, the optical power is fixed to $P=1 \, W$, whose relation with the photon-number distribution $n_m(\omega)$ is discussed in the Appendix~A. Note that, for the particular case of standard silica fibers, the above examples would correspond to $n_{wg}=1.4516, n_{0}=1.4316$ at $\lambda=2.4\,\mu m$ for ${T}>0$, and $n_{wg}=1.4723, n_{0}=1.4623$ at $\lambda=0.5\,\mu m$ for ${T}<0$, where $n_{wg}$ and $n_{0}$ are the refractive indices of the waveguide and the surrounding medium, respectively, with $V_0=\pi(n_{wg}^2-n_{0}^2)/(\lambda n_{wg})$.

In Fig.~\ref{fig:no_cond}(b)-(e), the temporal spectra are plotted as a function of $\sqrt{|\kappa|} \omega$ to show that the typical spectral widths scale as $\sim 1/\sqrt{|\kappa|}$. In this respect, it is important to note that, by an appropriate change of variables \cite{note_kappa}, the dispersion parameter can be rescaled out of the equilibrium distribution (\ref{eq:BE_to_RJ}), thereby eliminating $|\kappa|$ from all subsequent analysis presented in this work. Consequently, apart from its sign, the exact magnitude of the dispersion parameter $|\kappa|$ has no qualitative impact on the thermodynamic equilibrium properties of the system.


\section{Positive temperatures}
\label{sec:pos_temp}

\subsection{Spatial analysis in a step-index waveguide}
\label{subsec:Spatial_cond}

When considering the basic features of BE condensation in low-dimensional systems~\cite{stringari16}, we may recall that a phase transition to BE (or RJ) condensation can occur in a 2D parabolic shaped trapping potential, such as a multimode GRIN fiber \cite{zanaglia24}, but not in  a 2D homogeneous step-index waveguide. Here there is no phase transition to BE or RJ condensation in the thermodynamic limit and
any macroscopic population of the fundamental mode in a step-index waveguide stems from a finite size effect, which eventually disappears when the system size is increased (see, e.g., Fig.~5 in  \cite{zanaglia24}).

While these works have dealt with a two-dimensional spatial-only dynamics, in the following we address the issue of three-dimensional, spatio-temporal light condensation in a step-index waveguide in the full (2+1)D ST configuration. This investigation takes inspiration from a number of studies on the coherence properties of an atomic beam evaporatively cooled in a magnetic guide~\cite{Castin2000,DGO_th,DGO_exp,Raithel,Schreck}. In this context, it was shown that the atomic beam experiences two-dimensional (2D) transverse BE condensation, rather than a full 3D condensation. Because of the very elongated geometry of the magnetic guide~\cite{Castin2000}, the system was shown to exhibit a macroscopic population of the fundamental mode in the transverse 2D spatial dimensions, while preserving long-wavelength fluctuation that destroy long-range order in the longitudinal direction. A full condensation can only be recovered in the limit of a shallow transverse confinement.

Here, we revisit this problem in the context of multimode optical waveguides, where the transverse waveguide supports a finite number of spatial modes, i.e., the potential that confines the optical field in the transverse spatial dimension is truncated, $V_0 < \infty$. We first study the thermal equilibrium properties of the transverse spatial mode distribution, and then show that, by increasing the number of spatial modes $M$, the temporal spectrum exhibits significant narrowing, until full (2+1)D ST condensation is achieved in the thermodynamic limit.

\subsubsection{Finite number of spatial modes}

Let us recall that we consider a continuous temporal incoherent optical wave, characterized by fluctuations that are statistically stationary in  time. Then the relevant quantity for characterizing the field is not the number of photons $N$, but the longitudinal density of photons, $\rho=N/L$ (in m$^{-1}$). 
Following \cite{chiocchetta16}, the longitudinal photon density takes the form:
\begin{align}
\rho &= \sum_{m=0}^{M-1}   \int \frac{d \omega}{2\pi v_g}\frac{1}{e^{(\beta_m+|\kappa| \omega^2-{\mu})/{T}} -1} 
\label{eq:rho_beyond_TL_1} \\
&= \frac{\sqrt{\pi}}{2\pi v_g} \sqrt{\frac{T}{|\kappa|}} \sum_{m=0}^{M-1} 
g_{1/2}\Big( e^{({\mu}-\beta_m)/{T}} \Big) 
\label{eq:rho_beyond_TL}
\end{align}
where $g_p(z)=\frac{1}{\Gamma(p)}\int_0^\infty dx \frac{x^{p-1}}{z^{-1} e^x-1}=\sum_{l=1}^\infty \frac{z^l}{l^p}$ is the Bose function, and $\zeta(p)=g_p(1)$ is the Riemann function \cite{stringari16}.
In other terms, in (\ref{eq:rho_beyond_TL}) we have considered the continuous (thermodynamic) limit in the time domain, because there is no confinement along the temporal dimension; whereas we keep a discrete sum over the transverse spatial modes of the waveguide. 
We recall that the longitudinal photon density is related to the optical power of the beam by $\rho={P}/(v_g \hbar {\tilde \omega}_0)$.

Denoting by $\rho_0$ the photon density in the fundamental spatial mode $m=0$ (i.e., integrated along the frequency $\omega$), the fraction of photon density that populates the fundamental transverse mode is
\begin{align}
\frac{\rho_0}{\rho} = 1 -  \frac{\sqrt{\pi}}{2\pi v_g \rho} \sqrt{\frac{T}{|\kappa|}} \sum_{m\neq0}^{M-1} g_{1/2}\Big( e^{({\mu}-\beta_m)/{T}} \Big).
\label{eq:rho_0_beyond_TL}
\end{align}
Consequently, $\rho_0/\rho$ can be viewed, in a loose sense, as the `spatial condensate' fraction of the ST system for a finite transverse size of the waveguide, i.e., for finite number of modes $M$.

\begin{figure}
\includegraphics[width=1\columnwidth]{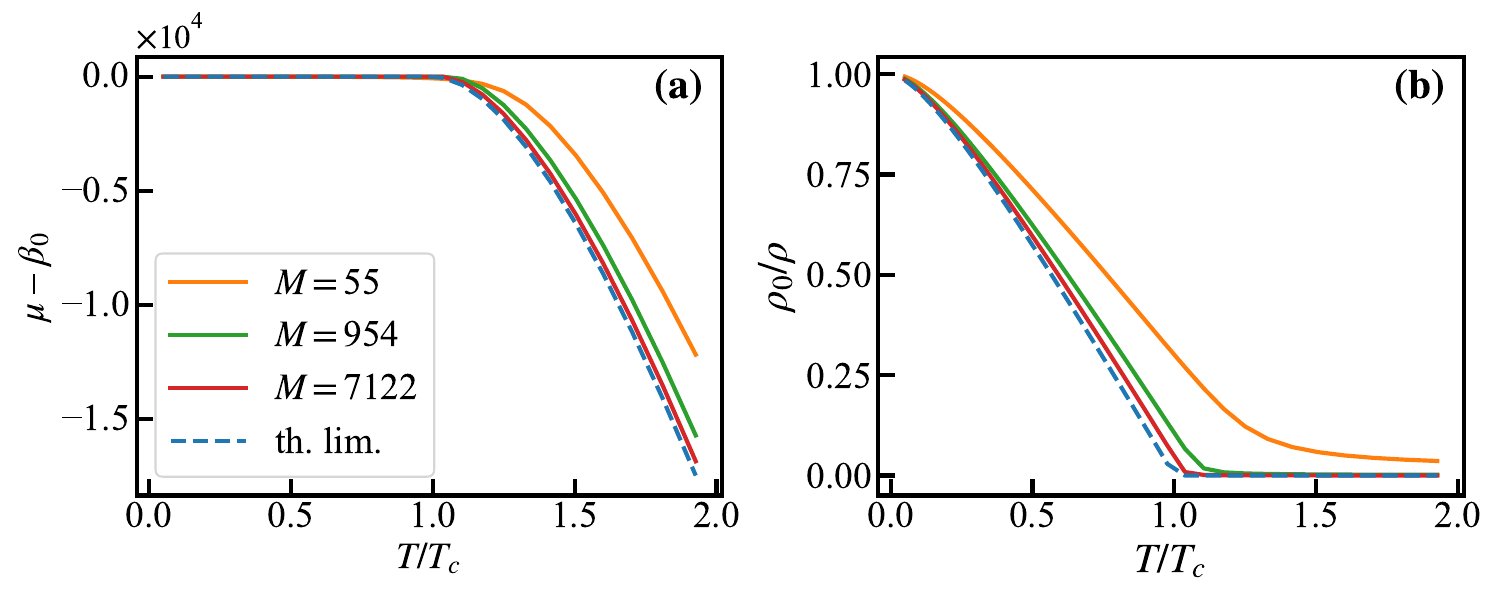}
\caption{
\baselineskip 10pt
{\bf Condensation at positive temperatures.} Convergence to the thermodynamic limit for ${T}>0$ in a step-index waveguide:
(a) chemical potential vs temperature, ${\mu}({T})$; 
(b) spatial condensate fraction vs temperature, $\rho_0({T})/\rho$.   
The solid lines refer to the computation of the discrete sums beyond the thermodynamic limit, from Eq.(\ref{eq:rho_beyond_TL}) for (a), and from Eq.(\ref{eq:rho_0_beyond_TL}) for (b). By increasing the number of modes $M$ (or the waveguide surface $S$) while keeping the photon density $\rho/S=$const and $V_0=$const, the curves approach the thermodynamic limit (dashed blue line), which are obtained from Eq.(\ref{eq:rho_TL_00}) for (a), and from Eq.(\ref{eq:frac_spat_cond_TL}) for (b). In the thermodynamic limit, ${\mu}({T})$ in (a), and $\rho_0({T})/\rho$ in (b), correspond to the (2+1)D ST condensation: both curves display a singular cusped behavior at ${T} = {T}_c$, where  ${\mu}=\beta_{0}$ for $0<{T} \le {T}_c$, and $\rho_0=0$ for ${T} > {T}_c$, where the critical temperature ${T}_c$ is given by Eq.(\ref{eq:T_c_pos}).
Parameters are given in the text.
}
\label{fig:cond_Tpos} 
\end{figure}
  
\subsubsection{Toward the continuous thermodynamic limit}

In the following, we consider highly multimode waveguides, and extend the procedure described in \cite{zanaglia24} for the pure 2D spatial case to the ST (2+1)D problem under consideration. More precisely, we study how the system approaches the thermodynamic limit, that is the limit where the waveguide surface tends to infinity $S \to \infty$, while keeping constant the three-dimensional photon density $\rho/S=N/(SL)=$const (in m$^{-3}$) and the potential depth, $V_0=$const \cite{zanaglia24}. Note that the photon density $\rho/S$ is proportional to the optical intensity $I=P/S$ because $\rho=P/(v_g \hbar {\tilde \omega}_0)$, where $P$ is the optical power. In this way, for a given beam intensity $I$ (or equivalently $\rho/S$), Eq.(\ref{eq:rho_beyond_TL}) provides a closed relation between the chemical potential and the temperature ${\mu}({T})$, which is reported in Fig.~\ref{fig:cond_Tpos} for different values of $S$ (i.e., different number of modes $M$). As evidenced in Fig.~\ref{fig:cond_Tpos}, as the system size $S$ increases (or equivalently $M$ increases), the curves ${\mu}({T})$ converge to the continuous limit (dashed blue line), where the chemical potential reaches the fundamental eigenvalue ${\mu} = \beta_0$, below a certain critical temperature ${T}_c$ (which will be defined below). In the example of Fig.~\ref{fig:cond_Tpos}, we considered a step-index waveguide with $V_0=52 \, 726\,m^{-1}$,  
$\kappa=-5.3\times 10^{-24}\,s^2\cdot m^{-1}$, 
while the number of modes $M$ (or the surface $S$) is increased by keeping constant the three-dimensional photon density $\rho/S=2.07\times 10^{16}\,m^{-3}$ (or intensity $I=P/S= 3.54\times 10^5\,W\cdot m^{-2}$).

Analyzing the thermodynamic limit is essential. It is in this limit that phase transitions are usually recognized and characterized by a singular (discontinuous) behavior of the thermodynamic quantities when the control parameter (e.g., the temperature) is varied across the transition point. In addition, the advantage of the continuous limit is that it enables us to derive thermodynamic relations in explicit analytical form. In the continuous approach, the discrete sum over the spatial modes can be converted to continuous integrals
\begin{equation}
\sum_{m=0}^{M-1} \to \int_0^{V_0} d\beta\, \varrho(\beta),    
\label{eq:cont_TDL}
\end{equation}
where $\varrho(\beta)$ is the density of states (DOS) \cite{stringari16}. The continuous approach is justified when the equilibrium distribution populates a large number of modes, that is, the optical temperature is much higher than the typical spacing between mode eigenvalues ${T} \gg \delta \beta$ \cite{stringari16}.
We recall that, as commented above through Eqs.(\ref{eq:rho_beyond_TL_1}-\ref{eq:rho_beyond_TL}), we have considered the continuous thermodynamic limit in the temporal dimension, along which there is no confinement.

Referring to the step-index waveguide under consideration in Fig.~\ref{fig:cond_Tpos}, the DOS does not depend on the propagation constant, $\varrho(\beta)=\varrho_0=k_0 S/(2\pi)$, as typical for 2D systems with no confinement potential. 
In the continuous thermodynamic limit, the longitudinal photon density is obtained from Eq.(\ref{eq:rho_beyond_TL}) by converting the discrete sum over the transverse $m$ modes into a continuous integral [see (\ref{eq:cont_TDL})], which gives
\begin{align}
\rho  = \frac{\varrho_0\sqrt{\pi}}{2\pi v_g} \, \frac{{T}^{3/2}}{\sqrt{|\kappa|} }   
\Big[ g_{3/2}(z) -g_{3/2}\big(z e^{-V_0/{T}}\big)  \Big],
\label{eq:rho_TL_00}
\end{align}
where $z=e^{({\mu}-\beta_0)/{T}}$ is the fugacity.
Note that the second term in the square brackets in (\ref{eq:rho_TL_00}) is specific to our waveguide geometry featured by a truncated potential, $V_0 < \infty$.  
The analysis of Eq.(\ref{eq:rho_TL_00}) reveals that, by decreasing the temperature, the chemical potential reaches the fundamental mode eigenvalue, ${\mu} \to \beta_0^-$ (i.e., $z \to 1^-$) for a non-vanishing critical temperature ${T}_c >0$, which is solution of 
\begin{align}
\rho = \frac{\varrho_0 \sqrt{\pi}}{2 \pi v_g} \, \frac{{T}_c^{3/2}}{\sqrt{|\kappa|} }  
\Big[ \zeta(3/2) - g_{3/2}\big(e^{-V_0/{T}_c}\big)  \Big].
\label{eq:T_c_pos}
\end{align}
For lower temperatures $T<{T}_c$, a macroscopic fraction $\rho_0/\rho$ of the density is concentrated in the $m=0$ transverse mode~\cite{Castin2000}. 
A plot of the chemical potential vs temperature is reported in Fig.~\ref{fig:cond_Tpos}(a): as the number of modes increases, the curves tend to the thermodynamic limit curve indicated by the dashed blue line.

An explicit formula for the spatial condensate fraction for ${T} \le {T}_c$ can be obtained by considering Eq.(\ref{eq:rho_0_beyond_TL}) in the continuous limit and setting ${\mu}=\beta_0$,
\begin{align}
\frac{\rho_0}{\rho} = 1 - \Big( \frac{{T}}{{T}_c} \Big)^{3/2} \ 
\frac{ \zeta(3/2) -g_{3/2}\big(e^{-V_0/{T}}\big)  }{ \zeta(3/2) -g_{3/2}\big(e^{-V_0/{T}_c}\big)    }.
\label{eq:frac_spat_cond_TL}
\end{align}
A plot of the spatial condensate fraction is reported in Fig.~\ref{fig:cond_Tpos}(b): once again, as the number $M$ of modes increases, the curve tends to the thermodynamic limit result \eqref{eq:frac_spat_cond_TL} indicated by the dashed blue line.

\subsection{ST condensation in a step-index waveguide}
\label{subsec:ST_cond}

Let us now analyze the phase transition to full ST condensation in the lowest $\omega=0$ state of the fundamental mode $m=0$. 
We follow the usual treatment of BE condensation in the thermodynamic limit \cite{stringari16}. For ${T} < {T}_c$, we split the contribution of the singularity of the BE distribution at $\omega=0$ and $m=0$, and the thermal contribution $\rho = \rho_{00}+ \int_0^{V_0} d\beta \varrho_0  \int \frac{d\omega}{2\pi v_g} \big( e^{(\beta+|\kappa|\omega^2)/{T}} -1  \big)^{-1}$ in all other states. 

According to the standard theory of BE condensation~\cite{stringari16}, the photon gas then undergoes a phase transition to ST condensation, characterized by a macroscopic population of the $\omega=0$ state of the fundamental spatial mode $m=0$,  
\begin{align}
\frac{\rho_{00}}{\rho} = 1 - \Big( \frac{{T}}{{T}_c} \Big)^{3/2} \ 
\frac{ \zeta(3/2) -g_{3/2}\big(e^{-V_0/{T}}\big)  }{ \zeta(3/2) -g_{3/2}\big(e^{-V_0/{T}_c}\big)    },
\label{eq:frac_cond_TL_00}
\end{align}
where the critical temperature ${T}_c$ is still given by Eq.(\ref{eq:T_c_pos}).
Accordingly, the ST condensate fraction $\rho_{00}/\rho$ vanishes for ${T} > {T}_c$, while it  increases to 1 when the temperature decreases below ${T}_c$ and tends to $0$. 

It is important to stress that, in the thermodynamical limit, the condensate fraction for the spatial case in Eq.(\ref{eq:frac_spat_cond_TL}), and the ST case in Eq.(\ref{eq:frac_cond_TL_00}), are identical: the total population $\rho_0$ in the fundamental spatial mode $m = 0$ (i.e., integrated along the frequency $\omega$) is in fact dominated by the population $\rho_{00}$ of the fundamental mode $m=0$ at $\omega=0$, while the contribution to the photon density at $m=0$ provided by the $\omega\neq 0$ states is negligible. This is expected in the continuous approximation in the ST domain. Indeed, the expression (\ref{eq:BE_to_RJ}) shows that only $m=0$, $\omega=0$ may present a singularity. Therefore, in the continuous approximation in the ST domain the photon distribution presents a singular (Dirac-$\delta$) contribution at $\beta=0$, $\omega=0$ with weight $\rho_{00}$ and a continuous contribution with  density with respect to the 2D Lebesgue measure in $\beta$ and $\omega$ and with total weight $\rho - \rho_{00}$. We remind the reader that the 2D Lebesgue measure of a line is zero and that any measure that has a density with respect to the 2D Lebesgue measure satisfies the same property (because any measure that admits a density with respect to the 2D Lebesgue measure automatically assigns measure zero to any set that has 2D Lebesgue measure zero). Thus, the integral with respect to $\omega$ at $\beta=0$ (that gives $\rho_0$) is equal to the weight $\rho_{00}$ of the singular contribution at $\beta=0$, $\omega=0$.

We finally remark that Eq.(\ref{eq:rho_TL_00}) encompasses the standard criterion for BE condensation of a dilute 3D Bose gas in the thermodynamic limit~\cite{stringari16}. Indeed, in the absence of a cut-off waveguide ($V_0 \to \infty$), we can follow \cite{chiocchetta16} by introducing the thermal wavelengths in the transverse and longitudinal dimensions, 
$\lambda_{\perp,\parallel} = \big( 2 \pi \hbar^2 /(m_{\perp,\parallel} k_B T) \big)^{1/2}= \big( 4 \pi \alpha_{\perp,\parallel} / {T} \big)^{1/2}$, with the effective masses $m_{\perp}=\hbar  k_0/v_g$, $m_\parallel=\hbar/(2v_g^3 |\kappa|)$, and corresponding spatial (diffraction) and temporal (dispersion) coefficients, $\alpha_\perp=1/(2k_0)$ and $\alpha_\parallel=\kappa v_g^2$, respectively. In this way, Eq.(\ref{eq:rho_TL_00}) can be written at ${T}={T}_c$ in the form $\lambda_\parallel \lambda_\perp^2 \rho/S=\zeta(3/2)$. 
This expression recovers the standard criterion for BE condensation in 3D: condensation arises at $\lambda_\parallel \lambda_\perp^2 \rho/S \sim 1$, which can be reached either by increasing the photon density $\rho/S$, or by decreasing the temperature ${T}$, i.e., by increasing the thermal wavelengths.

\begin{center}
\begin{figure}[!h]
\includegraphics[width=0.99\columnwidth]{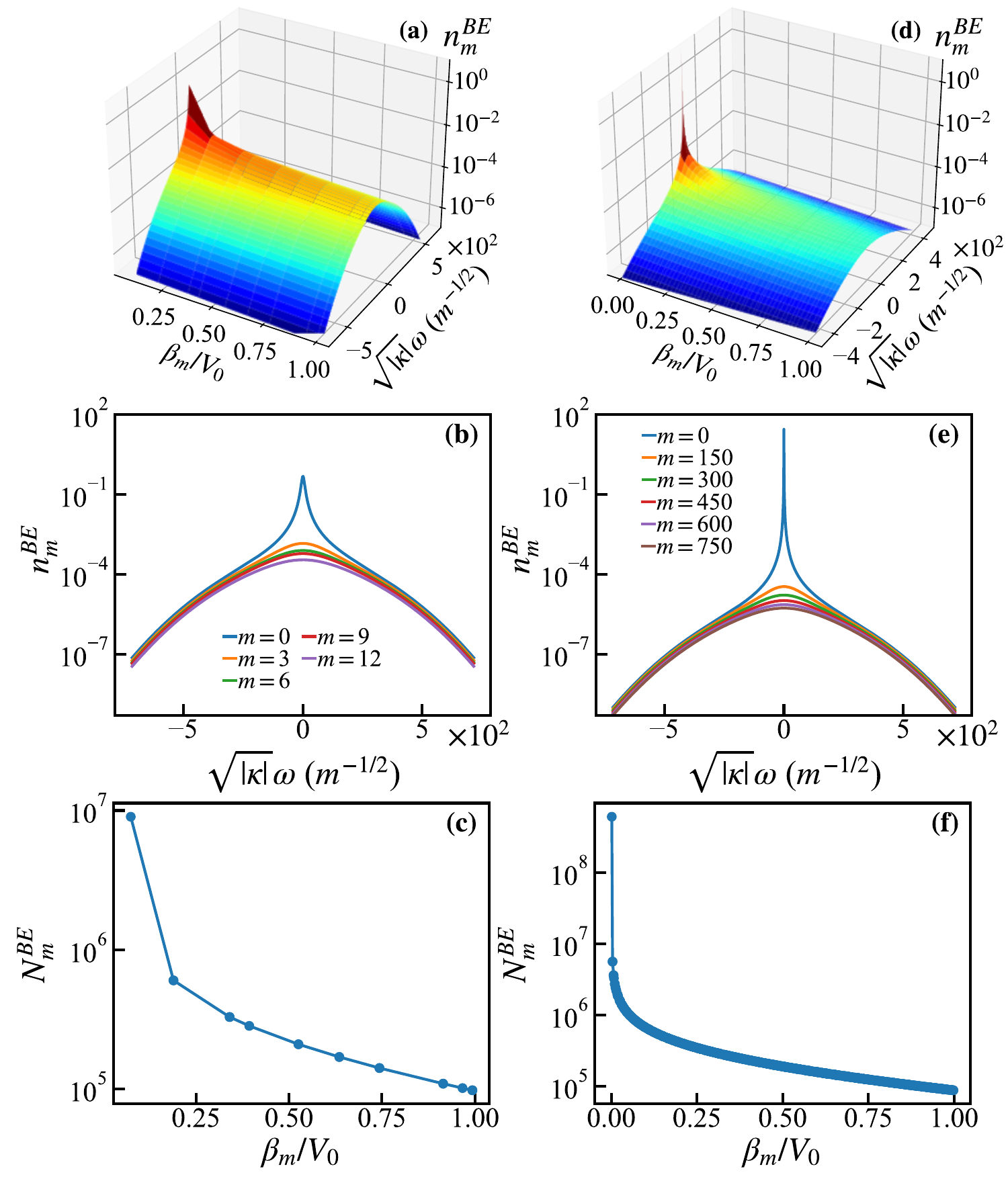}
\caption{
\baselineskip 10pt
{\bf Spatial mode distribution and temporal spectra in the condensed regime at positive temperatures.} Equilibrium distribution (\ref{eq:BE_to_RJ}) for a step-index waveguide at ${T}=0.6 {T}_c$ for $M=17$ [left column: (a-b-c)], and $M=954$ [right column: (d-e-f)]. 
The plots on the first row report the ST distribution $n_m^{BE}(\omega)$ from Eq.(\ref{eq:BE_to_RJ}), the plots on the second and third rows display cross-sections of the corresponding distributions in the top row: temporal spectra $n_m^{BE}(\omega)$ for different modes $m$ (2nd row);  spatial mode distribution integrated in frequency, ${N}_m^{BE}=(2\pi v_g)^{-1}\int n_m^{BE}(\omega) d\omega$ (3rd row). 
By increasing the number of modes $M$, we observe a significant spectral narrowing of the fundamental mode (compare (b) and (e)), as well as a macroscopic population of the fundamental spatial mode $m=0$ (compare (c) and (f)). Parameters are given in the text.
}
\label{fig:3Dplots_cond}
\end{figure}
\end{center}

\subsection{Spectral narrowing by increasing the number of modes}
\label{subsec:spect_narr_Tpos}

We have seen that 2D spatial condensation tends to evolve into full (2+1)D ST condensation by increasing the transverse surface size of the waveguide. This suggests that the temporal spectrum of the fundamental mode should exhibit significant spectral narrowing as the waveguide becomes wider. We illustrate this remarkable property in Fig.~\ref{fig:3Dplots_cond}, which reports the ST equilibrium distribution $n_m^{BE}(\omega)$ given in Eq.(\ref{eq:BE_to_RJ}) for a small waveguide ($M=17$ modes, left column), and a large waveguide ($M=954$ modes, right column). In this figure, we have considered a partially condensed case where ${T}=0.6 {T}_c$. Note that although ${T} < {T}_c$, we have ${\mu} \neq \beta_0$, because the waveguides have a finite size and the system is not in the thermodynamic limit. Accordingly, the value of the chemical potential ${\mu}$ is obtained by solving Eq.(\ref{eq:rho_beyond_TL}). In the example of Fig.~\ref{fig:3Dplots_cond} we have considered the following parameters: $\rho/S=2.1\times 10^{16}\,m^{-3}$, $P=0.1 \, W$, $V_0=52 \, 725\,m^{-1}$,  
with waveguide radius $R=14 \mu m$ for $M=17$, and $R=100 \mu m$ for $M=954$.   
 
The comparison of Fig.~\ref{fig:3Dplots_cond}(b) and Fig.~\ref{fig:3Dplots_cond}(e) reveals the occurrence of a significant spectral narrowing of the fundamental mode as the waveguide size (or number of modes $M$) increases. 
At the same time, the fundamental spatial mode gets macroscopically populated, as can be seen by the comparison of Fig.~\ref{fig:3Dplots_cond}(c) and Fig.~\ref{fig:3Dplots_cond}(f).

\subsection{BE vs RJ regimes}

The BE distribution is known to recover the classical RJ distribution in the limit of highly occupied modes. However, as discussed above in Sec.~\ref{subsec:uv_cat}, the RJ distribution leads to the well-known UV catastrophe (signaled, e.g., by a divergence of the energy $E$) in the absence of a high-frequency cut-off. Nonetheless, it is worth noting that the one-dimensional integral over temporal frequencies appearing in the expression of the photon density  (\ref{eq:rho_beyond_TL_1}) remains convergent for the RJ law. Based on this observation, we proceed to assess the validity of the RJ approximation by formally substituting the BE law with the RJ law, while keeping in mind that the RJ regime is prone to UV divergences in the ST context.

Estimating the number of photons $N$ in usual beam-cleaning experiments using optical pulses with a finite temporal duration is not immediate, because $N \simeq P \tau/(\hbar {\tilde \omega}_0)$ depends on the temporal duration $\tau$ of the (quasi-)monochromatic optical pulse. This would lead to the inconsistent conclusion that an experiment performed with a CW laser beam would involve an infinite number of photons. As previously mentioned, the natural relevant parameter to study (2+1)D ST thermalization is not the total number of photons, but the longitudinal photon density $\rho$. 

\begin{figure}
\includegraphics[width=1\columnwidth]{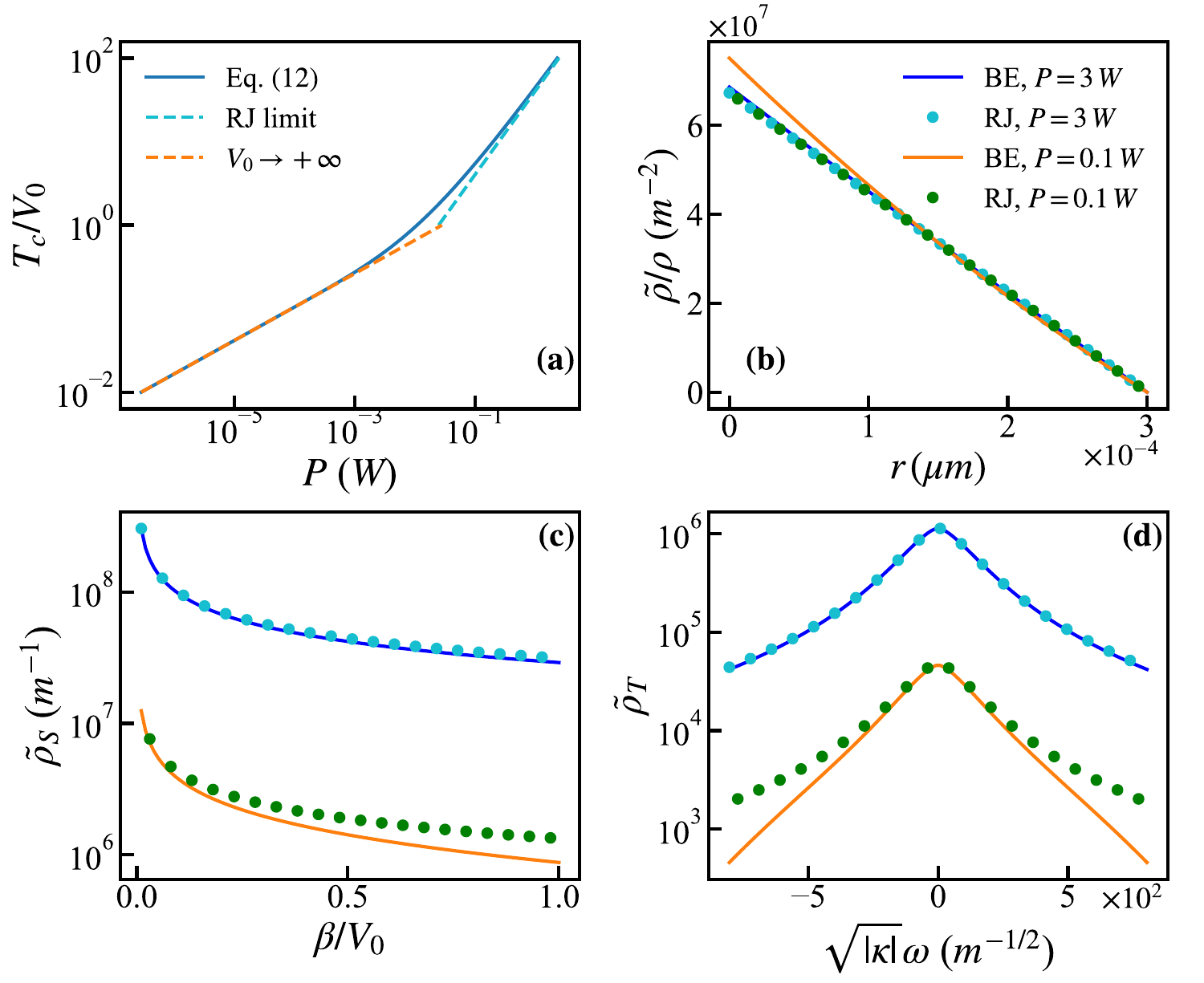}
\caption{\baselineskip 10pt
{\bf BE vs RJ regimes.} Relation between the optical power $P$ and the critical temperature ${T}_{c}$ to condensation, from Eq.(\ref{eq:rho_grin_Tpos}) (blue line), for a parabolic multimode waveguide. 
For large powers $P \gg 0.1$W, the photon density can be approximated by the RJ limit (dashed orange line, ${T}_c \sim P$). By decreasing the power, the photon density follows the BE expression (\ref{eq:rho_grin_Tpos}) with $V_0 \to \infty$ (dashed green line, ${T}_c \sim P^{2/5}$).
Transverse 2D intensity distribution ${\tilde \rho}(r)$ (with $r=|{\bm r}_\perp|$) (b), spatial mode distribution ${\tilde \rho}_S(\beta)$ (c), and temporal spectrum ${\tilde \rho}_T(\omega)$ (d), computed with the BE law (solid lines), and the corresponding RJ approximation (dotted lines), at ${T}={T}_c$. As seen in panels (b) to (d), at large power ($P=3 \, W$), the distributions are well approximated by the RJ law, while at small power ($P=0.1 \, W$) noticeable deviations emerge. 
Parameters are given in the text.
}
\label{fig:be_rj}
\end{figure}

We illustrate this in Fig.~\ref{fig:be_rj}(a), which reports the transition from the BE to the RJ regime by increasing the optical power, which is related to the photon density by $\rho=P/(v_g \hbar {\tilde \omega}_0)$, as previously discussed. In this example, we considered a parabolic waveguide 
(with radius $R=300\mu$m, $V_0=52 \, 726$m$^{-1}$, 
$\kappa=-5.3\times 10^{-24}\,s^2\cdot m^{-1}$).
The DOS for a truncated parabolic waveguide is $\varrho(\beta)=\beta/\beta_0^2$ for $\beta \le V_0$, and $\varrho(\beta)=0$ for $\beta > V_0$ \cite{aschieri11,zanaglia24}.
Accordingly, the photon density in the continuous limit at ${T}={T}_c$ reads  
\begin{align}
\rho = \frac{1}{2v_g \beta_0^2} \frac{{T}_c^{5/2}}{\sqrt{\pi |\kappa|}}  
\Big[ \zeta(\frac{5}{2}) -g_{5/2}\big(e^{- \frac{V_0}{{T}_c} }\big) 
- \frac{V_0}{{T}_c} g_{3/2}\big(e^{-\frac{V_0}{{T}_c}}\big) \Big].
\label{eq:rho_grin_Tpos}
\end{align}
For large powers, all modes are highly occupied, and the photon density is well approximated by the RJ limit $\rho^{RJ}  =\frac{V_0^{3/2}}{3v_g \beta_0^2 \sqrt{|\kappa|}} {T}_c$, see the dashed-orange line in Fig.~\ref{fig:be_rj}(a). By decreasing the power, quantum effects associated to the exponential decay of the BE distribution at high energy play a role, and 
the photon density follows the behavior $\rho \simeq  \frac{\zeta(5/2)}{2v_g \beta_0^2 \sqrt{\pi |\kappa|}}{T}_c^{5/2}$, which is obtained from Eq.(\ref{eq:rho_grin_Tpos}) without spatial frequency cutoff ($V_0 \to \infty$), see the dashed green line in Fig.~\ref{fig:be_rj}(a). More precisely, in the example of Fig.~\ref{fig:be_rj}, we considered a parabolic waveguide with $G=94$ groups of degenerate modes, where the number of guided modes is $M=G(G+1)/2 \simeq G^2/2$ for $G \gg 1$. 

\subsubsection{Spatial mode distribution and temporal spectra}

The validity of the RJ approximation can be assessed through the analysis of the spatial mode distribution ${\tilde \rho}_S(\beta)$, and the temporal spectrum ${\tilde \rho}_T(\omega)$ of the field, defined as the two marginal distributions of $n(\beta,\omega)$,
\begin{align}
&{\tilde \rho}_S(\beta)= \int \frac{d \omega}{2\pi v_g} n^{BE}(\beta,\omega),
\label{eq:rho_s_beta}\\
&{\tilde \rho}_T(\omega)= \int_0^{V_0} d\beta \varrho(\beta) n^{BE}(\beta,\omega).
\label{eq:rho_t_w}
\end{align}
We report in Fig.~\ref{fig:be_rj}(c)-(d) plots of such marginal distributions at ${T}={T}_c$, which  are computed with the BE law $n(\beta,\omega) = \big(e^{(\beta+|\kappa| \omega^2)/{T}_c}- 1\big)^{-1}$ (solid lines), and the corresponding RJ approximation $n^{RJ}(\beta,\omega)={T}_c/(\beta+|\kappa| \omega^2)$ (dotted lines). 
We note in panels (c)–(d) of Fig.~\ref{fig:be_rj} that for high power, the distributions closely follow the RJ approximation, whereas for low powers a clear deviation is observed.

\subsubsection{Spatial intensity distribution}

The measurement of the modal distribution in an optical fiber experiment is known to be quite complicated, see, e.g.,  \cite{mangini22,pourbeyram22,Gervaziev20,kibler23}. Here, we discuss another important quantity that is experimentally accessible and that can be exploited to discriminate between the classical and quantum regimes. This quantity refers to the spatial intensity distribution of the optical field in the transverse spatial dimension ${\bm r}_\perp$. It can be calculated from the semi-classical approximation by writing the equilibrium distribution in the form \cite{stringari16}:
\begin{align}
n^{BE}({\bm k}_\perp,{\bm r}_\perp,\omega)= \Big( e^{\frac{\alpha_{\perp} k_{\perp}^{2} + |\kappa| \omega^{2}+V({\bm r}_\perp)-{\mu}}{{T}}}-1 \Big)^{-1} ,
\label{eq:n_eq_SC}
\end{align}
where $V({\bm r}_\perp)$ is the waveguide trapping potential in the transverse dimension, ${\bm k}_\perp$ the corresponding wave-vector, and recall that $\alpha_\perp=1/(2k_0)$ is the diffraction coefficient. The spatial density distribution is obtained by integration in both spatial and temporal frequency space, which gives 
\begin{align}
{\tilde \rho}({\bm r}_\perp) &= \int \frac{d^2{\bm k}_\perp}{(2\pi)^2} \int \frac{d \omega}{2\pi v_g} n^{BE}({\bm k}_\perp,{\bm r}_\perp,\omega) \nonumber \\
&=\frac{1}{8 \pi^{3/2}v_g}\,\frac{{T}^{3/2}}{\alpha_\perp\sqrt{|\kappa|}}  
\Big[ g_{3/2}\Big( e^{-\frac{V({\bm r}_\perp)-{\mu}}{{T}}}\Big) 
\nonumber \\
& \quad - g_{3/2}\Big( e^{-\frac{V_0-{\mu}}{{T}}}\Big) \Big].
\label{eq:int_dist_grin}
\end{align}
The corresponding intensity distribution in the RJ approximation reads ${\tilde \rho}^{RJ}({\bm r}_\perp)=\frac{{T}}{4\pi v_g \alpha_{\perp} \sqrt{|\kappa|}} \big(\sqrt{V_0-{\mu}}-\sqrt{V({\bm r}_\perp)-{\mu}}\big)$, which is obtained from $n^{RJ}({\bm k}_\perp,{\bm r}_\perp,\omega)=  {T}/\big(\alpha_{\perp} k_{\perp}^{2} + |\kappa| \omega^{2}+V({\bm r}_\perp)-{\mu}\big)$. Considering the example of a parabolic waveguide, we report in Fig.~\ref{fig:be_rj}(b) the intensity distribution (solid lines), and the corresponding RJ approximation (dotted lines), at ${T}={T}_c$. As for panels (c)-(d), a good agreement is obtained with the RJ approximation at high power. Conversely, at low power a pronounced, cusp-like deviation from the RJ distribution appears near the center of the trapping potential.

We finally note that, by integrating the density distribution ${\tilde \rho}({\bm r}_\perp)$ in the 2D transverse spatial domain over $|{\bm r}_\perp|\le R$ ($R$ being the waveguide radius with $V(|{\bm r}_\perp|=R)=V_0$), we recover the total density of photons discussed above in the continuous limit
\begin{align}
\rho=\int_{|{\bm r}_\perp|\le R}  {\tilde \rho}({\bm r}_\perp)d{\bm r}_\perp=\int \frac{d \omega}{2\pi v_g} \int_0^{V_0} d\beta \varrho(\beta)  n^{BE}(\beta, \omega).
\end{align}

\section{Bose-Einstein condensation at negative temperatures}
\label{sec:BEC_Tneg_grin}

The idea of negative temperatures was originally proposed in the seminal works by Onsager \cite{onsager49} and Ramsey \cite{ramsey56}. Significant efforts have been devoted to the theoretical understanding of these unusual thermodynamic equilibrium states. 
The concept of negative temperatures is now broadly accepted, as it has been the subject of different experimental observations \cite{frenkel15,buonsante16,puglisi17,cerino15,baldovin21,onorato21,skipp21}. 
As a matter of fact, negative temperatures were originally experimentally observed in the context of nuclear spin systems \cite{purcell51}, and more recently with cold atoms in optical lattices \cite{braun13}, and in 2D quantum superfluids \cite{gauthier19,johnstone19}. 
More recently, negative temperatures have been predicted for classical multimode optical wave systems in Ref. \cite{christodoulides19}, and subsequently observed in these systems with light waves in Refs.~\cite{PRL23,muniz23,ferraro25}. In particular, in the experiments reported in Ref.\cite{ferraro25}, negative temperatures were associated to the strongly nonlinear regime of a dense photon gas. 

As anticipated above, negative temperatures emerge as natural equilibrium states in ST thermalization, whenever one considers the normal dispersion regime, see Fig.~\ref{fig:no_cond}. We will now show that negative temperature ST equilibrium states exhibit a phase transition to BE condensation at $\beta=\beta_{\rm max}$ and $\omega=0$. We recall that in the purely 2D spatial domain,  BE condensation at negative temperatures does not exist, as discussed in \cite{zanaglia24}.  

\begin{center}
\begin{figure}
\includegraphics[width=1\columnwidth]{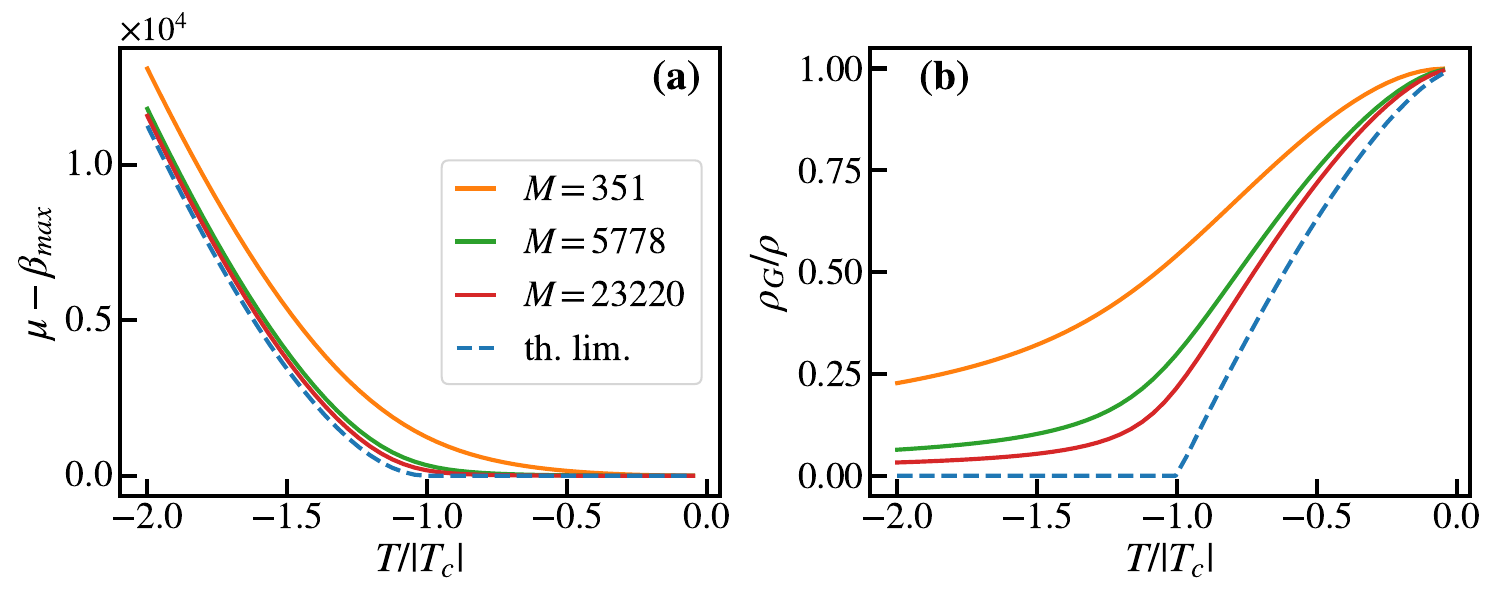}
\caption{
\baselineskip 10pt
{\bf Condensation at negative temperatures.} 
Convergence to the thermodynamic limit for ${T}<0$ in a parabolic waveguide: 
(a) chemical potential vs temperature, ${\mu}({T})$; 
(b) condensate fraction in the highest mode group vs temperature, $\rho_G({T})/\rho$.   
The solid lines refer to the computation of the discrete sums beyond the thermodynamic limit, from Eq.(\ref{eq:rho_beyond_TL_neg}) (a), and from Eq.(\ref{eq:spat_rho_0_beyond_TL}) (b). By increasing the number of modes $M$ (or the waveguide surface $S$) while keeping the photon density $\rho/S=$const and $V_0=$const, the curves approach the thermodynamic limit (dashed blue lines), which are obtained from Eq.(\ref{eq:rho_gen_Tc_neg}) for (a), and from Eq.(\ref{eq:gen_cond_fract_neg}) for (b). In the thermodynamic limit, ${\mu}({T})$ in (a), and $\rho_{G}({T})/\rho$ in (b), correspond to the (2+1)D ST condensation: both curves display a singular cusped behavior at ${T} = {T}_c <0$, where ${\mu}=\beta_{\rm max}=V_0$ for ${T}_c \le {T} \le 0$, and $\rho_G=0$ for ${T} < {T}_c <0$. 
Parameters are given in the text.
}
\label{fig:rho_beyond_TL_T_neg}
\end{figure}
\end{center}

\subsection{Spatial analysis}

ST condensation at negative temperatures can take place in a parabolic multimode waveguide or in a homogeneous step-index waveguide, as discussed in the Appendix~\ref{sec:appendix_STcond}. 
In the following, we consider the case of a parabolic waveguide (e.g., GRIN multimode fiber), because the highest energy level is highly degenerate, which makes the transition to BE condensation of a different nature with respect to conventional condensation in the (non-degenerate) fundamental mode for ${T}>0$. We recall that the mode eigenvalues read $\beta_m = \beta_0(m_x+m_y+1)$, where $m=(m_x,m_y)$ labels the two integers that specify a mode \cite{beta_m_note}, and that the waveguide supports $G$ groups of degenerate modes, with $M \simeq G^2/2$ the number of modes ($G \gg 1$).

Considering the normal dispersion regime $\kappa > 0$ (${T}<0$ and ${\mu} > V_0$), the photon density integrated along the frequency $\omega$ reads
\begin{align}
\rho = \frac{\sqrt{\pi}}{2\pi v_g} \sqrt{\frac{-\tilde T}{\kappa}} \sum_{m} 
g_{1/2}\Big( e^{(\beta_m-{\mu})/(-{T})} \Big) .
\label{eq:rho_beyond_TL_neg}
\end{align}
Denoting by $\rho_G$ the photon density that populates the highest transverse modes with eigenvalue  $\beta=\beta_{\rm max}=V_0$, we have 
\begin{align}
\frac{\rho_G}{\rho} = 1 -  \frac{\sqrt{\pi}}{2\pi v_g \rho} \sqrt{\frac{-\tilde T}{\kappa}} \sum_{m} {\vphantom{\sum}}' g_{1/2}\Big( e^{(\beta_m-{\mu})/(-{T})} \Big),
\label{eq:spat_rho_0_beyond_TL}
\end{align}
where $\sum_{m}^{'}$ denotes a sum over all modes except the highest mode group $\beta_{\rm max}$. 

We now study how the system approaches the thermodynamic limit by following the procedure of section~\ref{subsec:Spatial_cond}. By keeping fixed the beam intensity $I=P/S$ (or equivalently the photon density $\rho/S=$const), Eq.(\ref{eq:rho_beyond_TL_neg}) provides a closed relation between ${\mu}$ and ${T}$, which is reported in Fig.~\ref{fig:rho_beyond_TL_T_neg} for different values of the surface $S$ (i.e., different $G$), while keeping constant $\rho/S$ and $V_0$. 
Note that the waveguide surface is $S=\pi R^2$, where we recall that the waveguide radius is defined by $V(|{\bm r}_\perp|=R)=V_0$. Consequently, the surface scales as $S \sim V_0/\beta_0^2$, and the thermodynamic limit for the parabolic waveguide is taken by keeping fixed $\rho \beta_0^2=$const and $V_0=$const. 
In the example of Fig.~\ref{fig:rho_beyond_TL_T_neg}, we considered a parabolic waveguide with $V_0=126 \,093 \,m^{-1}$,  
$\kappa=3.6\times 10^{-25}\,s^2\cdot m^{-1}$,  
with $\rho/S=4.4\times 10^{16}\,m^{-3}$ (or $I=P/S= 3.5\times 10^6\,W\cdot m^{-2}$). We note in Fig.~\ref{fig:rho_beyond_TL_T_neg}(a) that, as the transverse waveguide surface $S$ increases, the curves ${\mu}({T})$ converge to the continuous limit, which will be discussed in the next section. 

\subsection{ST condensation in the highest energy level}

Following the analysis of section~\ref{sec:pos_temp}, the spatial condensate fraction of the photon density (\ref{eq:spat_rho_0_beyond_TL}) converges to the ST condensate fraction by increasing the size of the waveguide toward the thermodynamic limit. We now analyze the phase transition to ST condensation in the highest mode group $\beta_{\rm max}$ at $\omega=0$. 
In a way akin to the case of positive temperatures, condensation at negative temperatures originates in the singularity of the BE equilibrium distribution. By increasing the negative temperature, the chemical potential decreases, and reaches the highest energy level ${\mu} = \beta_{\rm max}=V_0$ at a non-vanishing negative critical temperature, ${T}_c <0$ (see Fig.~\ref{fig:rho_beyond_TL_T_neg}(a)). Considering the continuous limit, the longitudinal photon density reads $\rho=\int \frac{d \omega}{2\pi v_g} \int_0^{V_0} d\beta \varrho(\beta)  n(\beta, \omega)$, where we recall that the DOS for a parabolic waveguide is given by $\varrho(\beta)=\beta/\beta_0^2$ for $\beta \le V_0$, and $\varrho(\beta)=0$ for $\beta > V_0$. At the critical temperature ${T}={T}_c$, we set ${\mu}=V_0$, and the integration  over $\beta$ gives
\begin{align}
\rho 
=&\frac{V_0^2}{2\pi v_g\beta_0^2} \int_{-\infty}^{\infty} d\omega
\Big[ \frac{{T}_c}{V_0} \ln\big(1- e^{\frac{\kappa \omega^2}{{T}_c}}  \big) - \frac{{T}_c^2}{V_0^2} g_2\big( e^{\frac{\kappa \omega^2}{{T}_c}}  \big)
\nonumber\\
&+  \frac{{T}_c^2}{V_0^2} g_2\big( e^{\frac{\kappa \omega^2+V_0}{{T}_c}}  \big)\Big].
\end{align}

\begin{center}
\begin{figure}[!h]
\includegraphics[width=1\columnwidth]{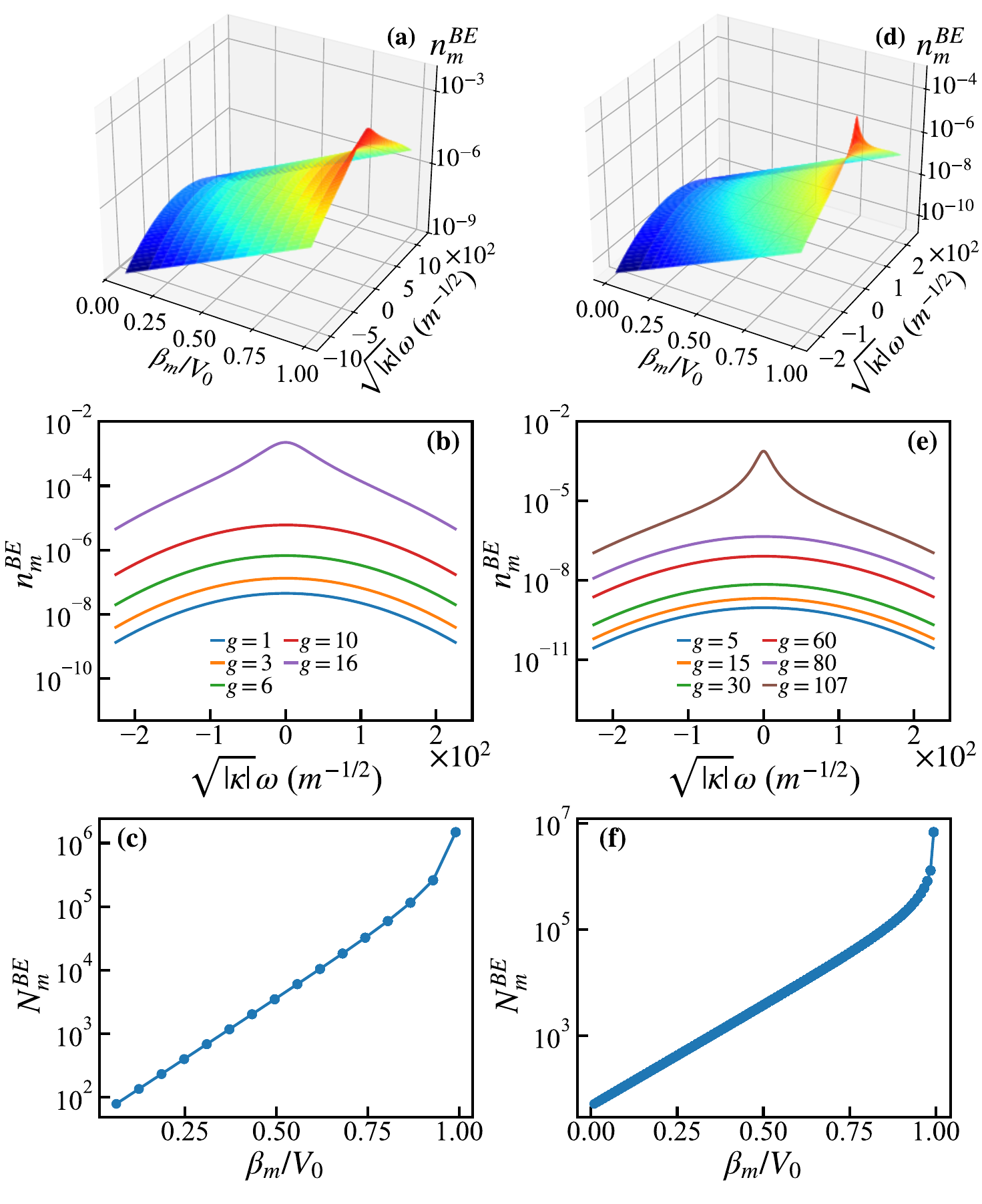}
\caption{
\baselineskip 10pt
{\bf Spatial mode distribution and temporal spectra in the condensed regime at negative temperatures.} 
Equilibrium distribution (\ref{eq:BE_to_RJ}) for a parabolic multimode waveguide (e.g., GRIN mutimode fiber) in the normal dispersion regime ($\kappa>0$) leading to negative temperatures ${T}=- 0.4 |{T}_c|$:  for $G=16$ groups of degenerate modes ($M=136$ modes) see the left column, and for $G=107$ groups of degenerate modes ($M=5 \, 778$ modes)  see the right column. 
The plots on the first row report the ST distribution $n_m^{BE}(\omega)$ from Eq.(\ref{eq:BE_to_RJ}), the plots on the second and third rows display cross-sections of the corresponding distributions in the top row: temporal spectra $n_m^{BE}(\omega)$ for different mode groups $g$ (2nd row);  spatial mode distribution integrated in frequency, ${N}_m^{BE}=(2\pi v_g)^{-1}\int n_m^{BE}(\omega) d\omega$ (3rd row). 
In (b) the curves refer (from top-to-bottom) to the mode group $g=1, 3, 6, 10, 16$; while in (e) they refer to the mode group $g=5, 15, 30, 60, 80, 107$. 
Remark the significant spectral narrowing in the highest mode group (top curve in (b) and (e)) when the number of modes increases.
Parameters are given in the text.
}
\label{fig:3Dplots}
\end{figure}
\end{center}

We remark that the first term of the integrand diverges for a vanishing dispersion coefficient, $\kappa=0$. 
This reflects the fact that there is no BE condensation in the pure 2D spatial domain for negative temperatures \cite{zanaglia24}. 
Here, thanks to the longitudinal temporal dimension, the photon density $\rho$ converges, which defines a non-vanishing negative critical temperature ${T}_c$ for inverted BE condensation:   
\begin{align}
\rho = \frac{|{T}_c|^{5/2}}{2v_g\sqrt{\pi \kappa} \beta_0^2} 
\Big[ - \frac{V_0}{{T}_c}  \zeta(\frac{3}{2}) - \zeta(\frac{5}{2}) 
+  g_{5/2} \big(  e^{ \frac{V_0}{{T}_c}}\big) \Big].
\label{eq:Tc_neg}
\end{align}
For ${T}>|{T}_c|$ and ${\mu}=V_0$, the singularity of the BE distribution is regularized by a macroscopic population of the highest energy level $\beta_{\rm max}$ at the frequency $\omega=0$. The fraction of particle density condensed in the highest energy level $\beta_{\rm max}=V_0$ at $\omega=0$ is 
\begin{align}
\frac{\rho_{G0}}{\rho} = 1 - \Big( \frac{{T}}{{T}_c} \Big)^{5/2} \ 
\frac{ g_{5/2}\big(e^{V_0/{T}}\big) -\zeta(5/2)  - \frac{V_0}{T} \zeta(3/2) }{ g_{5/2}\big(e^{V_0/{T}_c}\big) -\zeta(5/2)  - \frac{V_0}{{T}_c} \zeta(3/2)   }.
\label{eq:gen_cond_fract_neg}
\end{align}
We recall that $\rho_{G0}$ denotes the total population of the $G$-fold degenerate energy level associated with the  eigenvalue $\beta_{\rm max}$ (at $\omega=0$), so that the corresponding population of each individual degenerate mode is reduced by a factor $G$. 
According to (\ref{eq:gen_cond_fract_neg}), the condensate vanishes at the negative critical temperature, $\rho_{G0}=0$ at ${T}={T}_c <0$, and the condensate fraction increases toward one as the temperature increases to zero, $\rho_{G0}/\rho \to 1$ as ${T} \to 0^-$. 
As compared to \eqref{eq:frac_cond_TL_00}, note that the different value $5/2$ of the exponent (instead of $3/2$) in (\ref{eq:gen_cond_fract_neg}) and the presence of different $g_p(z)$ functions is due to the parabolic waveguide geometry under consideration (the case of a homogeneous step-index waveguide is considered in the Appendix~\ref{sec:appendix_STcond}).

Using the same argument as in section~\ref{subsec:ST_cond}, $\rho_{G0}$ is also the particle density in the highest energy level $\beta_{\rm max}$ integrated along the frequency $\omega$, that is $\rho_{G}$. This is because the contribution of the density at $\beta_{\rm max}$ and $\omega\neq 0$ is negligible in the ST continuum limit. Accordingly, in the thermodynamic limit, the ST condensate given by Eq.(\ref{eq:gen_cond_fract_neg}) coincides with the spatial condensate that populates the highest energy level
\begin{align}
\rho_{G0}({T})=\rho_{G}({T}).
\label{eq:rhoG0_rhoG}
\end{align}
We finally note that, for ${T} < {T}_c < 0$,
the photon density reads 
\begin{align}
\rho =& \frac{|{T}|^{5/2}}{2v_g \beta_0^2 \sqrt{\pi\kappa}}  
\Big[ - \frac{V_0}{T} g_{3/2}\big(e^{({\mu}-V_0)/{T}}\big) -g_{5/2}\big(e^{({\mu}-V_0)/{T}}\big) \nonumber \\
&+g_{5/2}\big(e^{{\mu}/{T}}\big) \Big].
\label{eq:rho_gen_Tc_neg}
\end{align}
As expected, this expression well recovers Eq.(\ref{eq:Tc_neg}) when the critical point is approached, that is in the limit ${T} \to {T}_c^-$ and ${\mu} \to V_0^+$.

\subsection{Spectral narrowing by increasing the number of modes}
\label{subsec:spect_narr_Tneg}

We have seen that 2D spatial condensation tends to evolve into a fully (2+1)D ST condensation by increasing the transverse size of the waveguide. This suggests that also in the negative temperature case, the temporal spectrum of the highest energy level should exhibit significant spectral narrowing as the waveguide becomes wider. We illustrate this property in Fig.~\ref{fig:3Dplots}, which reports the ST equilibrium distribution $n_m^{BE}(\omega)$ given in Eq.(\ref{eq:BE_to_RJ}) for a parabolic multimode waveguide with $G=17$ (left column), and $G=108$ (right column), groups of degenerate modes. Here, we have considered a partially condensed case where ${T}=-0.4 |{T}_c|$, while the corresponding value of the chemical potential is obtained by solving Eq.(\ref{eq:rho_beyond_TL_neg}) for a given value of $\rho/S$. We considered in Fig.~\ref{fig:3Dplots} the following parameters: $V_0=126 \,093 \,m^{-1}$,  
$P=1 \, W$, i.e.,  $\rho/S=4.4\times 10^{16}\,m^{-3}$ or $I=P/S= 3.5\times 10^6\,W\cdot m^{-2}$. We remark that spectral narrowing in Fig.~\ref{fig:3Dplots} occurs for the highest mode group at $\beta=\beta_{\rm max}$. Note that the less pronounced peak around $\omega=0$ in (b,e) compared to the corresponding plots in Fig.~\ref{fig:3Dplots_cond} is related to the strong degeneracy of the highest-order mode for the parabolic waveguide under consideration.

\section{Conclusions}
In this work, we have developed a unified theoretical framework of spatio-temporal (ST) thermalization of light propagating in multimode optical fibers. On the classical side, extensive numerical simulations of the nonlinear Schr\"odinger (NLS) equation demonstrate thermalization to a ST Rayleigh–Jeans (RJ) equilibrium state in agreement with the corresponding wave turbulence kinetic equation. Peculiar adiabatic cooling phenomena associated to the UV behaviour of classical RJ thermalization are pointed out and the consequent limitations of the classical approximation are highlighted.

To go beyond and properly include quantum effects, we start from the general theory of light propagation in nonlinear media to derive a quantum NLS equation. This is used to show that the corresponding kinetic equation governs relaxation toward a Bose-Einstein (BE) equilibrium distribution with a fully regular UV behavior. All together, this analysis provides a general understanding of the ST equilibrium properties of optical waves propagating in multimode waveguides.

In analogy with atomic beams evaporatively cooled in a magnetic guide~\cite{Castin2000,DGO_th,DGO_exp,Raithel,Schreck}, our approach led us to predict that ST incoherent light waves exhibit a transition to condensation when the transverse size (i.e., the number of transverse modes) of the waveguide is increased for a fixed power per unit surface, eventually leading to complete (2+1)D ST condensation in the thermodynamic limit. The transition to condensation is characterized by different properties depending on the dispersion regime of the waveguide.

For anomalous dispersion, the condensed state is characterized by a macroscopic occupation of the fundamental spatial mode, accompanied by a dramatic spectral narrowing of light carried in the same mode. Conversely, in the normal dispersion regime the ST equilibrium distribution is characterized by negative temperatures, whereby high-order spatial modes are more populated than low-order modes, while the temporal spectrum remains peaked at the fundamental (carrier) frequency. In this way, ST incoherent light exhibits an inverted transition to BE condensation at negative temperatures, which occurs by increasing the temperature above a negative critical value. The ST condensed state is characterized by a macroscopic population of the highest energy level $\beta=\beta_{\rm max}$, accompanied by a dramatic spectral narrowing of light for that propagation constant.
In addition, we have shown that the ST configuration allows one to discriminate the classical RJ equilibrium from the quantum BE equilibrium: this can be done through the analysis of the spatial mode distribution and temporal spectrum, but also through the transverse spatial near-field intensity distribution, a quantity that can easily be measured in an optical experiment. 


Work is in progress to extend the present analysis to nontrivial ST coupling by considering a general model of light propagation that goes beyond the paraxial and slowly-varying-envelope approximations. Along this line, ST thermalization can be investigated by considering temporally incoherent (ASE) sources, which may exploit ST instabilities \cite{wrightPRL15,krupa16} or supercontinuum generation \cite{wright16,krupa16b,gentyNC22} in multimode fibers. The observation of ST thermalization requires minimizing the detrimental impact of the dissipative Raman effect~\cite{PRL08,PRA22}: as an alternative to optical fibers, other experimental platforms may be considered to suppress this effect, such as optically induced waveguides in atomic vapors \cite{glorieux22,glorieux23} photorefractive crystals \cite{denz03}, or gas-filled hollow-core photonic crystal fibers \cite{russell14}.

In the recent work \cite{zanaglia25}, it has been shown that the temporal degrees of freedom can dramatically accelerate the rate of thermalization to thermal equilibrium. On the other hand, different studies based on wave turbulence theory~\cite{PRL19,PRA19,PRL22,ferraro25_rev} have revealed that a significant acceleration of the nonequilibrium process of thermalization in multimode fibers is also provided by the structural disorder and the random mode coupling due to refractive index fluctuations introduced by inherent imperfections and environmental perturbations \cite{mecozzi12a,mecozzi12b,mumtaz13,xiao14,psaltis16,caramazza19}.
Since the aforementioned studies \cite{PRL19,PRA19,PRL22,ferraro25_rev} have been carried out within the framework of purely spatial dynamics, 
a natural and important extension will be to incorporate structural disorder into the ST wave turbulence theory developed in Ref.\cite{zanaglia25}. This would allow for a more comprehensive understanding of how disorder influences the nonequilibrium dynamics of ST thermalization in multimode optical systems.

It would also be interesting to extend the present study to investigate the onset of condensation in regimes beyond the weakly nonlinear limit, where linear dispersion effects no longer dominate over the nonlinear Kerr contribution. The stronger nonlinearities in this regime holds the potential to reveal a wealth of new physical phenomena. In this regime, the condensate no longer simply coexists with the uncondensed thermal background, but instead emerges as a distinct phase characterized by remarkable physical behaviors that are absent in the weakly nonlinear case. For instance, its superfluidity properties and the emergence of collective Bogoliubov sound waves~\cite{michel18,fontaine18,michel21} may have a significant impact on the thermalization process. Moreover, the dynamics can become highly complex and turbulent, exhibiting phenomena such as the formation and interaction of quantized vortices, i.e., quantum turbulence \cite{Tsatsos16}. The prospect of investigating ST superfluidity, and ST quantum turbulence at positive, or even negative, temperatures within an optical multimode waveguide represents a compelling and largely unexplored frontier, open for future research.

We finally recall that we have considered in this work the purely conservative propagation geometry of the optical beam through a Kerr nonlinear material. It would be interesting to extend the notion of ST  thermalization to cavity configurations, where non-equilibrium processes of condensation have been studied~\cite{bloch22}, either in the genuine quantum BE regime \cite{kasprzak06,weitz10,weitz10b,weitz16,nyman18,barland21,fischer19,fischer21} or in the classical regime \cite{conti08,PRA11b,berloff13,weill10,turitsyn12,turitsyn15,fratalocchi15,PRA15}.
Extending the analysis to optical cavities could therefore open qualitatively new perspectives, allowing one to investigate how thermalization mechanisms are reshaped by the interplay of driving, dissipation, and confinement, and to probe novel regimes of non-equilibrium optical self-organization.


\section{Acknowledgement}
Funding was provided by Agence Nationale de la Recherche (Grants No. ANR-23-CE30-0021, No. ANR-21-ESRE-0040). Calculations were performed using HPC resources from Universit\'e C\^ote d’Azur’s Center Azzura, and DNUM-CCUB (Universit\'e de Bourgogne Europe). I.C. acknowledges financial support from the PE0000023-NQSTI project by the Italian Ministry of University and Research, co-funded by the European Union - NextGeneration EU; from Provincia Autonoma di Trento (PAT), partly via the Q@TN initiative. M.F. acknowledges funding from the Italian Ministry of University and Research (MUR) through the Italian Science Fund (FIS2), project SOFT (H53C24001530001). S.W. acknowledges support from the European Innovation Council, project MULTISCOPE (101185664).

\bigskip

\appendix
\section{Relation between the classical wave distribution and the photon-number distribution\label{app:A} }

In this Appendix, we derive the relation between the distribution ${\tilde n}_m(\omega)$ discussed  in the classical wave limit through the kinetic Eq.(\ref{eq:kin_class}) [see Sec.~\ref{sec:ST_eq}], and the photon-number distribution $n_m(\omega)$ used in the quantum kinetic formalism of Eq.(\ref{eq:kin}) [see Sec.~\ref{sec:quantum}].

In the classical wave limit described by the NLS Eq.(\ref{eq:nls_rt}), the optical power reads
\begin{align}
{P} =  
 \int d\br_\perp \left< |\psi|^2 \right> = \frac{1}{(2\pi)^2} \sum_{m=0}^{M-1} \int  {\tilde n}_m(\omega) d\omega,
\label{eq:P_app}
\end{align}
where  the factor $1/(2\pi)^2$ comes from the adopted Fourier transform convention, see Eq.(\ref{eq:Four}).

On the other hand, the (temporally constant) optical power can be related to the homogeneous 
longitudinal photon density along the spatial direction $\zeta$ by 
\begin{align}
\rho={P}/(v_g \hbar {\tilde \omega}_0),
\label{eq:rho_P_rel_app}
\end{align} 
where ${\tilde \omega}_0$ is the carrier optical frequency. The longitudinal photon density can be related by 
\begin{align}
\rho= \sum_{m=0}^{M-1}   \int \frac{d k_\zeta}{2\pi} 
n_m(k_\zeta),
\label{eq:rho_app}
\end{align}
to the $k_\zeta$-space photon-number distribution $n_m(\zeta)$ in the mode $m$ per unit effective length $\zeta=v_g t$ at effective conjugate momentum $k_\zeta=\omega/v_g$ defined via
\begin{equation}
\langle {\phihd}_m(k_\zeta)\,{\phih}_n(k'_\zeta)\rangle= 2\pi\,\delta_{mn}\,\delta(k_\zeta-k'_\zeta)\,n_m(k_\zeta)
\end{equation}
where $\phih_m(k_\zeta)$ is the Fourier transform 
\begin{equation}
\phih_n(k_\zeta)=\int\,d\zeta\,e^{-ik_\zeta \zeta}\,\phih_n(\zeta)
\end{equation} 
of the renormalized field operator $\phih_m(\zeta)=\psih_m(\zeta)/\sqrt{\hbar\tilde{\omega}_0v_g}$ satisfying standard Bose commutation rules,
\begin{equation}
[\phih_m(\zeta;\tau),\phihd_n(\zeta';\tau)]=
\delta_{mn}\,
\delta(\zeta-\zeta')
\end{equation}
and thus referring to the photon number.
Note that $n_m(\omega)$ can be equivalently understood as  the frequency-space photon-number distribution $n_m(\omega)$ in the mode $m$ per unit time. 

By comparing Eqs.~(\ref{eq:P_app}) and (\ref{eq:rho_app}), we obtain from (\ref{eq:rho_P_rel_app}) the following relation between the distribution ${\tilde n}_m(\omega)$ for classical waves and the quantum photon-number distribution $n_m(\omega)$:
\begin{align}
n_m(\omega) = {\tilde n}_m(\omega) / (2\pi \hbar {\tilde \omega}_0).
\label{eq:n_app}
\end{align}
Note that this relation is also valid out of equilibrium, i.e., the modal populations $n_m(\omega)$ and ${\tilde n}_m(\omega)$ do not necessarily refer to the equilibrium distributions.




\section{Influence of first-order dispersion \label{app:B}}

The analysis presented in Sec.~\ref{sec:ST_eq} is based on the general form of the ST NLS Eq.(\ref{eq:nls_rt}), which implicitly neglects the group-velocity mismatch between the different modal components (the modal components $b_m(\omega)$ in Eq.(\ref{eq:nls}) propagate with the same group-velocity). In this Appendix, we demonstrate that explicitly accounting for the group-velocity difference does not modify the conclusions of Sec.~\ref{sec:ST_eq}, as discussed  through Eqs.~(\ref{eq:sign_kappa1}–\ref{eq:sign_kappa2}).

The group-velocity difference modifies the modal dispersion relation in the form ${\tilde \beta}_m(\omega) = \beta_m - w_m \omega - \kappa \omega^2$, where $w_m$ denote the inverse group-velocities of the modal components at the carrier frequency ${\tilde \omega}_0$. The group-velocity difference leads to a spectral shift of the modal components. The derivation of the equilibrium distribution therefore requires imposing the constraint of momentum conservation ${{\cal M}}=\frac{1}{(2\pi)^2} \sum_m \int \omega {\tilde n}_m(\omega) d\omega$ Ref.\cite{zanaglia25} (see Sec.~II of the Supplementary Material), in addition to the energy $E=\frac{1}{(2\pi)^2} \sum_m \int {\tilde \beta}_m(\omega) {\tilde n}_m(\omega) d\omega$ and the power ${P}=\frac{1}{(2\pi)^2} \sum_m \int {\tilde n}_m(\omega) d\omega$. Maximizing the entropy under the constraints imposed by the conservation of $({E},{P},{\cal M})$ then leads to a generalized form of the ST RJ distribution \cite{zanaglia25}:
\begin{align}
{\tilde n}_m^{RJ}(\omega) = \frac{\tilde T}{{\tilde \beta}_m(\omega) + \lambda \omega - {\tilde \mu}},
\label{eq:n_rj_momentum}
\end{align}
where $\lambda$ is the Lagrange multiplier associated to the conservation of the momentum ${\cal M}$. 
Considering an initial condition characterized by ${\cal M}=0$ (e.g., all spectra ${\tilde n}_m(\omega,z=0)$ are symmetric and centered at $\omega=0$), gives 
\begin{align}
\lambda = \sum_m w_m {P}_m^{eq}/{P}
\label{eq:lambda}
\end{align}
where ${P}_m^{eq}=\frac{1}{(2\pi)^2} \int {\tilde n}_m^{RJ}(\omega) d\omega$ is the power in the mode $m$ at equilibrium. The parameter $\lambda$ then denotes an average inverse group-velocity of the field \cite{pitois06,lagrange07}. Note in particular that neglecting first order dispersion $w_m=0$, then $\lambda=0$, and the ST equilibrium distribution (\ref{eq:n_rj_momentum}) recovers the distribution (\ref{eq:n_rj_general}) discussed in Sec.~\ref{sec:ST_eq}.

The RJ distribution (\ref{eq:n_rj_momentum}) can be written in the following Lorentzian form
\begin{align}
{\tilde n}_m^{RJ}(\omega) = \frac{-{\tilde T}/\kappa}{  \big(\omega - \frac{\lambda - w_m}{2 \kappa} \big)^2 - 
\big( \frac{\lambda - w_m}{2 \kappa} \big)^2 - \frac{\beta_m - {\tilde \mu}}{\kappa} }.
\label{eq:n_rj_mom_s}
\end{align}
It becomes apparent that the physical condition ${\tilde n}_m^{RJ}(\omega) >0$ imposes the following constraints on the thermodynamic parameters:
\begin{align}
&\kappa < 0 \quad \to \quad  {\tilde T}> 0, \ {\tilde \mu} < \beta_m -(\lambda - w_m)^2/(4|\kappa|) ,
\label{eq:sign_kappa1gvm}\\
&\kappa > 0 \quad \to \quad   {\tilde T}< 0, \ {\tilde \mu} > \beta_m + (\lambda - w_m)^2/(4|\kappa|).
\label{eq:sign_kappa2gvm}
\end{align}
The constraints on the thermodynamic parameters (${\tilde \mu}, \lambda$) then depend on the specific values of the modal group velocities $w_m$. In practice, however, one generally has $(\lambda - w_m)^2/(4|\kappa|) \ll \beta_m$. This can be seen by considering the nonlinear envelope equation, which provides a more general description of light propagation beyond the NLS equation, in particular by including modal group-velocity matching. Within this framework, the group velocity scales as $w_m \sim \beta_m/{\tilde \omega}_0$, where we recall that ${\tilde \omega}_0$ is the central carrier frequency. According to Eq. (\ref{eq:lambda}), the Lagrange multiplier scales as $\lambda \sim w_m$. As a consequence, the last term in Eqs. (\ref{eq:sign_kappa1gvm})–(\ref{eq:sign_kappa2gvm}) may be neglected when $\beta_m \ll |\kappa| {\tilde \omega}_0^2$, a condition that is well satisfied in standard multimode optical fibers. Under this assumption, the constraints on the thermodynamic parameters in Eqs. (\ref{eq:sign_kappa1gvm}–\ref{eq:sign_kappa2gvm}) reduce to those obtained in the absence of first-order dispersion (\ref{eq:sign_kappa1}-\ref{eq:sign_kappa2}). The conclusions then remain unchanged: in the anomalous-dispersion regime, the spatial equilibrium distribution is peaked on the fundamental mode at positive temperature, whereas in the normal-dispersion regime the distribution is inverted and corresponds to negative temperatures. We finally note that the analysis can be extended by considering the impact of high-order dispersion properties \cite{zanaglia25} (also see \cite{michel10,barviau13}).
\\

\section{Bose-Einstein condensation at negative temperatures in a step-index multimode waveguide}
\label{sec:appendix_STcond}

To complete our discussion in Sec.~\ref{sec:BEC_Tneg_grin}, we show that BE condensation at negative temperatures also occurs in step-index multimode waveguides. We consider Eq.(\ref{eq:rho_beyond_TL_neg}) in the continuous thermodynamic limit [see (\ref{eq:cont_TDL})], with the DOS  $\varrho(\beta)=\varrho_0=k_0 S/(2\pi)$. The longitudinal photon density $\rho=\int \frac{d \omega}{2\pi v_g} \int_0^{V_0} d\beta\, \varrho(\beta)\,  n(\beta, \omega)$ reads 
\begin{align}
\rho  = \frac{\varrho_0 (-{T})^{3/2}}{2v_g\sqrt{\pi \kappa}}    
\Big[  g_{3/2}\big(e^{({\mu}-V_0)/{T}}\big) -g_{3/2}\big(e^{{\mu}/{T}}\big) \Big],
\end{align}
where we recall that $\kappa>0$ and ${\mu} \ge V_0=\beta_{\rm max}$. 
We remark that ${\mu} \to V_0^+$ for a non-vanishing negative critical temperature ${T}_c$: 
\begin{align}
\rho  = \frac{\varrho_0 (-{T}_c)^{3/2}}{2v_g\sqrt{\pi  \kappa}}    
\Big[ \zeta(3/2) - g_{3/2}\big(e^{V_0/{T}_c}\big)   \Big],
\end{align}
To provide a complementary perspective to the previous parabolic waveguide analysis in Sec.~\ref{sec:BEC_Tneg_grin} (where the highest energy level was degenerate), here we consider a step-index waveguide with a non-degenerate highest energy level. By increasing the negative temperature above ${T}_c$, the photon gas undergoes a transition to BE condensation characterized by a macroscopic population of the highest mode $M$ with propagation constant  $\beta_{\rm max}=V_0$ at $\omega=0$:
\begin{align}
\frac{\rho_{M0}}{\rho} = 1 - \Big( \frac{{T}}{{T}_c} \Big)^{3/2} \ 
\frac{ \zeta(3/2) -g_{3/2}\big(e^{V_0/{T}}\big)   }{ \zeta(3/2) -g_{3/2}\big(e^{V_0/{T}_c}\big) }.
\end{align}
The condensate vanishes at and below the negative critical temperature, $\rho_{M0}/\rho=0$ at ${T}\leq {T}_c <0$, and increases to one as the temperature increases to zero, $\rho_{M0}/\rho \to 1$ as ${T} \to 0^-$.



\begin{thebibliography}{99}
\bibitem{beta_m_note}
For a circular step-index waveguide of radius $R$, $\beta_{m}=x_{l,s}^2/(2 k_0 R^2)$, where $m=(l,s)$ labels the two integers that specify a mode, with $x_{l,s}$ the $s$-th zero of the Bessel function of the first kind $J_l(x)$, and $k_0$ the light wave-number. For a parabolic (GRIN) multimode waveguide, $\beta_m = \beta_0(m_x+m_y+1)$, where $m=(m_x,m_y)$ labels the two integers that specify a mode, and $\beta_0$ is the fundamental mode eigenvalue.
\bibitem{note_kappa}
The dispersion parameter can be rescaled in the equilibrium distribution by means of the change of variables: ${\hat \beta}={\beta}/|\kappa|, {\hat T}={T}/|\kappa|, {\hat \mu}={\mu}/|\kappa|$. The corresponding rescaled density of states is given by ${\hat \varrho}({\hat \beta})={\varrho}({\beta})|\kappa|$, the diffraction parameter ${\hat \alpha}_\perp={\alpha}_\perp/|\kappa|$ and the potential depth ${\hat V}_0=V_0/|\kappa|$.
\end{thebibliography}

\end{document}